\documentclass[11pt,reqno]{article}

\usepackage{amsmath,amssymb,amsfonts,amsthm,bbm,mathrsfs,verbatim} 
\usepackage{booktabs}
\usepackage{setspace}
\usepackage{authblk}
\usepackage{caption}
\usepackage{geometry}
\usepackage{graphics} 
\usepackage{graphicx}  
\usepackage{multirow}
\usepackage{float}
\usepackage[ruled,vlined]{algorithm2e}
\usepackage{hyperref}
\usepackage{natbib}
\usepackage{color}
\usepackage{bm}

\newtheorem{theorem}{Theorem}
\newtheorem{proposition}{Proposition}

\newtheorem{lemma}{Lemma}
\theoremstyle{definition}

\newtheorem{definition}{Definition}

\def\R{{\mathbb R}}
\def\N{{\mathbb N}}
\def\E{{\mathbb E}}

\def\M{{\mathcal M}}

\def\SJ{{\Delta_J}}

\def\w{{\bf w}}

\allowdisplaybreaks

\definecolor{dblue}{RGB}{12, 56, 100}

\title{\bf Barycentric model aggregation in the Wasserstein space of distributions and a variational approach to consistency}

\author[a]{E. Androulakis}
\author[a]{G. I. Papayiannis}
\author[b]{A. N. Yannacopoulos}

\affil[a]{\footnotesize Department of Statistics \& Insurance Science, University of Piraeus, Piraeus, GR}	
\affil[b]{\footnotesize Department of Statistics, Athens University of Economics and Business, Athens, GR}	

\begin{document}
	\graphicspath{ {figures/} } 
	\maketitle

\begin{abstract}
	We study the problem of model aggregation within the Wasserstein space for probability measures on the real line. 
	Given a fixed finite collection of candidate probability models, we consider the associated class of Wasserstein barycenters and develop a data-driven calibration framework in which the aggregation weights are statistically learned from empirical information associated with a target distribution. From a variational perspective based on $\Gamma$-convergence, we establish consistency of the resulting aggregation scheme, showing that empirical minimizers converge to the minimizers of the actual problem, along with the associated barycentric estimators, under mild conditions. The performance of the proposed method is evaluated through synthetic experiments and illustrated on a real dataset from a temperature monitoring network of sensors.
\end{abstract}
	

\section{Introduction}\label{sec-1}

Modern statistical analysis increasingly involves data objects that are most naturally represented as probability distributions rather than as scalars or vectors. Such distribution-valued observations arise in a wide range of fields, including climate science and environmental modelling \citep{magyar2023hydrological, koundouri2024probabilistic, le2025barycenter}, economics and risk assessment \citep{avanzi2024ensemble, agosto2025new, spelta2025modelling, spelta2026density}, machine learning \citep{murphy2022probabilistic, pegoraro2022projected}, as well as in simulation-based inference \citep{sisson2018handbook, cranmer2020frontier} and Bayesian inference where posterior distributions constitute the primary objects of analysis (see e.g. \cite{backhoff2022bayesian}). In these settings, the full distribution captures essential, structural and uncertainty information that cannot be adequately summarized by low-dimensional statistics \citep{panaretos2019statistical, panaretos2020invitation}. Moreover, in many of the aforementioned settings, several probabilistic models, estimators or information sources are available for the same underlying quantity, with each one reflecting different modelling assumptions and spatial or temporal regimes. Combining these candidates into a single representative model is therefore a fundamental statistical task. This paper develops a data-calibrated barycentric aggregation framework for distributions in the Wasserstein space where the aggregation weights are not prescribed a priori but are statistically learned from empirical information associated with a target distribution, while the resulting calibration procedure and induced barycentric estimator are shown to be asymptotically consistent.

The Wasserstein barycenter provides a natural geometric notion of averaging probability measures, extending the notion of Euclidean mean to the space of distributions \citep{agueh2011barycenters, le2017existence}. Given a collection of probability models and a vector of non-negative weights summing to one, the corresponding barycenter is obtained as the distribution that minimizes a weighted sum of squared Wasserstein distances from the reference models. In this way, the aggregate distribution respects the geometry of probability measures and, in particular, accounts for the location, scale and shape features of the distributions being combined. While the classical Wasserstein barycenter framework typically relies on prescribed aggregation weights, calibration-based barycentric aggregation approaches have also been considered in more data-driven settings (see, e.g., \cite{papayiannis2018model, schmitz2018wasserstein}. In the present work, we consider a fixed finite collection of candidate probability models and study the statistical calibration of the aggregation weights from empirical information associated with a target distribution under the Wasserstein metric. The resulting procedure yields an empirically calibrated Wasserstein barycenter that provides an optimal aggregate representation of the target distribution within the barycentric class generated by the candidate models. Our main focus is on the asymptotic analysis of this calibration framework and the consistency properties of the induced barycentric estimator.

Our setting differs from the Wasserstein and Fr\'echet regression frameworks (see for instance \cite{petersen2019frechet, chen2023wasserstein, zhou2024wasserstein}) that appear in the literature. Within such regression frameworks, distribution-valued predictors can be employed as features (solely or combined with other types of covariates) when the response is distribution-valued. However, although there exist some conceptual similarities, these settings are entirely different in structure and purpose from the aggregation framework discussed in this work. Within the framework of barycentric model aggregation, the included models in the ensemble are interpreted as alternative probabilistic representations, measurements, or estimators of the same underlying quantity, namely the target distribution that needs to be approximated. These models may be provided by different agents, and are assumed to exhibit only moderate deviations from the target, rather than serving as covariates on a regression setting. Accordingly, our objective is to combine the available candidate distributions into a single aggregate model. This perspective differs fundamentally from distributional regression approaches, whose goal is to estimate distributional profiles conditional on explanatory variables. Moreover, beyond this conceptual distinction, the applicability of regression-based approaches in our setting is further limited by their typically substantial data requirements.

The one-dimensional Wasserstein framework considered here is particularly convenient and statistically interpretable. For the considered probability measures on the real line, the quadratic Wasserstein distance admits an equivalent representation in terms of the $L_2$ distance between quantile functions. Consequently, the Wasserstein barycenter associated with the weighting vector admits an explicit quantile representation as the corresponding convex combination of the quantile functions of the involved measures. This reduces the barycenter calibration problem to a standard (quadratic) Wasserstein cost minimization problem over the aggregation weights. The empirical version of this problem is obtained by replacing the target and candidate quantile functions by their empirical counterparts.

The main theoretical contribution of the paper is to establish the asymptotic validity of the proposed barycentric calibration procedure from a variational perspective, including both the convergence of the empirically calibrated aggregation weights and the induced Wasserstein barycenter. The existing literature on Wasserstein barycenters largely focuses on the consistency of empirical barycenters viewed as Fr\'echet means of random probability measures, typically in asymptotic regimes where the number of observed measures increases \citep{alvarez2016fixed, le2017existence,  bigot2018characterization, bigot2018upper}. In contrast, we consider a fixed, finite collection of candidate distributions and study the asymptotic behavior of a data-driven aggregation scheme as the sample size grows. Our analysis therefore concerns not only the barycenter itself, but the full calibration procedure, including the optimization over aggregation weights and the resulting barycentric estimator. In particular, employing the rigorous framework of $\Gamma$-convergence, we show that empirical minimizers converge to their population counterparts under mild conditions. Thus, the proposed method is asymptotically justified as an appropriately calibrated model aggregation procedure in the Wasserstein space.

The empirical performance of the proposed barycentric estimator is examined through simulation studies based on Weibull, Gamma and mixed distributional families. These synthetic-data experiments assess both the accuracy of the barycenter and the recovery of the aggregation weights under different sample sizes, different cardinalities of the set of available models and distributional configurations. We also perform a real-data application to spatial monitoring of maximum temperature distributions in the Thessaly region of Greece, that further demonstrates the practical utility of the approach, yielding stable predictions and revealing meaningful spatial connectivity patterns across monitoring sites.

The remainder of the paper is organized as follows: Section \ref{sec-2} introduces the barycentric model aggregation framework in Wasserstein space. Section \ref{sec-3} establishes the variational consistency results based on the notion of $\Gamma$-convergence. Section \ref{sec-4} presents the synthetic-data experiments, while Section \ref{sec-5} reports the real-data application to maximum temperature distributions monitoring. Section \ref{sec-6} concludes with a discussion of the main findings.

\section{The model aggregation setup in Wasserstein space}\label{sec-2}

Let us denote by $(\Omega, \mathcal{F}, \mathbb{P})$ a probability space. In this work we consider random variables where their induced probability measures have finite second moments, i.e. we consider probability measures in the space
$$ \mathcal{P}_2(\R) := \left\{ \mu \in \mathcal{P}(\R)  \,\, : \,\, \int_{\R} |x|^2 d\mu < \infty \right\} $$with $\mathcal{P}(\R)$ denoting the space of probability distributions in $\R$. For two random variables $X \sim \mu$, $Y \sim \nu$ where $\mu, \nu \in \mathcal{P}_2(\Omega)$, the (quadratic) Wasserstein distance between $\mu, \nu$ is defined as
\begin{equation*}
	\mathcal{W}_2(\mu, \nu) = \left\{ \inf_{\pi \in \Pi(\mu, \nu)} \E_{\pi}\big[ |X - Y|^2 \big] \right\}^{1/2}
\end{equation*}
where $\Pi(\mu,\nu)$ denotes the set of joint probability measures $\pi \in \mathcal{P}(\R \times \R)$ whose marginals coincide with the measures $\mu$ and $\nu$ (set of transport plans connecting $\mu$ and $\nu$). This metric arises from the theory of Optimal Transport and provides a proper metrization of the space of probability measures, as it is compatible with the topology of this space \citep{villani2009optimal, santambrogio2015optimal, villani2021topics}. In the special case of measures in $\R$ that we consider in this work, the Wasserstein distance admits the equivalent representation
\begin{equation*}
	\mathcal{W}_2(\mu, \nu) = \left\{ \E\bigg[ \big( q_{\mu}(U) - q_{\nu}(U) \big)^2 \bigg] \right\}^{1/2}, \quad U \sim \mathrm{Unif}(0,1)
\end{equation*}
where $q_{\mu}(\cdot), \, q_{\nu}(\cdot)$ denote the corresponding quantile functions of the probability measures $\mu, \nu$, respectively.

We consider the problem of constructing an estimator for a target distribution $\mu_0 \in \mathcal{P}_2(\R)$ within the Wasserstein space through an appropriate barycentric scheme. Assume that we have a collection of models 
$$ \mathcal{M} := \{ \mu_1, \mu_2, ..., \mu_J \} \subset \mathcal{P}_2(\R)$$
which can be realized as a finite ensemble of candidate models that will be utilized for the construction of an estimator for the target distribution $\mu_0$. To this end, we employ the notion of Fr\'echet mean \citep{frechet1948elements} on the set $\M$, which under the metric sense of the Wasserstein distance leads to the concept of Wasserstein barycenter. In this context, the induced barycentric estimator relies on a set of weights, which are appropriately calibrated based on the affinity of each model in $\M$ from the empirical evidence concerning the target model $\mu_0$. Let us denote by 
\begin{equation}\label{unit-simplex}
	\SJ :=  \bigg\{ \w = (w_1,w_2,...,w_J)' \,\, :\,\, \sum_{j=1}^J w_j=1, \,\,\, w_j \geq 0\quad  j=1,2,...,J \bigg\}
\end{equation}
the unit simplex, and assume that $\w\in \SJ$ represents the weighting vector. The Wasserstein barycenter of the set $\M$ for a given weighting vector $\w \in \SJ$ is determined as the minimizer
\begin{eqnarray}\label{wass-bar}
	\mu_B(\w) := \arg\min_{\mu \in \mathcal{P}(\mathcal{X})} \sum_{j=1}^J w_j \mathcal{W}_2^2(\mu, \mu_j).	
\end{eqnarray}
Existence and uniqueness results concerning problem \eqref{wass-bar} have been provided in the general setting of probability models by \cite{agueh2011barycenters, le2017existence} and in the manifold setting by \cite{kim2017wasserstein}. In the one-dimensional setting, the barycenter admits an explicit quantile representation, and in particular the Wasserstein barycenter $\mu_B(\w)$ stated in \eqref{wass-bar} can be written explicitly as
\begin{equation}\label{QA}
	q_B(s;\w) = \sum_{j=1}^J w_j \, q_j(s), \quad s \in (0,1)
\end{equation}
where $q_j(\cdot)$ denotes the quantile function of the probability measure $\mu_j$ for $j=1,2,...,J$. Note that since $w_j \geq 0$ for all $j=1,2,...,J$, the resulting function $q_B(\cdot;\w)$ remains a quantile function for any choice $\w \in \SJ$, since the relevant monotonicity property is preserved (i.e. for $s_1 < s_2$ where $s_1,s_2 \in (0,1)$ it holds that $q_B(s_1;\w) \leq q_B(s_2;\w)$ for any $\w\in\SJ$).

The resulting model in \eqref{QA}, referred to as the barycentric model or the Wasserstein barycenter, is the aggregate model that combines the information provided by the set $\M$, i.e. the one that attains simultaneously the minimal discrepancy by all models in $\M$ in the sense of the Wasserstein distance. Since the credibility of each model is represented by the relevant component in the weighting vector $\w$, an important issue is its optimal choice that corresponds to the optimal calibration problem of the barycenter $\mu_B(\w)$ to the empirical information from $\mu_0$. Assuming that a fair amount of empirical evidence have been collected from the target distribution $\mu_0$ (which is considered unknown), an empirical estimate $\mu_{0,n}$ is derived (from a sample of size $n$), and this estimation is incorporated for the calibration task of the barycentric model. In this perspective, a natural principle for the derivation of the optimal weights $\w$ is through the Wasserstein distance minimization problem (between $\mu_B(\w)$ and $\mu_{0,n}$), i.e.
\begin{eqnarray}\label{opt-w}
	\min_{\w \in \SJ} \frac{1}{2}\E\bigg[ \big( q_{B}(U;\w) - q_{0,n}(U) \big)^2 \bigg], \quad U \sim \mathrm{Unif}(0,1)
\end{eqnarray}
where $q_{0,n}(\cdot)$ denotes the empirical quantile function of $\mu_0$. Equivalently, the barycenter calibration problem can be written as the constrained optimization problem
\begin{eqnarray}\label{np}
	\min_{\w \in \SJ} \left\{ F(\w) := \frac{1}{2}\E\bigg[ \big( q_B(U;\w) - q_{0,n}(U) \big)^2  \bigg] \,\, : \,\, q_{B}(U;\w) = \sum_{j=1}^J w_j \, q_j(U) \right\}.
\end{eqnarray}
Existence and uniqueness results for a family of problems related to \eqref{np} are provided in \cite{papayiannis2018learning}.

The formulation in \eqref{np} emphasizes that the proposed barycentric model defines an estimator for the target distribution $\mu_0$, consistent with the Wasserstein geometry of the space of probability distributions, obtained through the minimization of the data-driven objective functional  ${\bf w} \mapsto F({\bf w})$ (see \eqref{np}) over  appropriate choice of the weights ${\bf w} \in \SJ$. This objective function (population objective) quantifies the discrepancy between the barycenter $\mu_B(\w)$ and $\mu_0$, while its empirical counterpart (i.e. the empirical objective induced by the sample) provides a stochastic approximation based on the induced empirical measure $\mu_{0,n}$ and the respective empirical quantile function $q_{0,n}(\cdot)$. The main issue in this approach is whether this empirical optimization problem constitutes a consistent approximation of the actual problem and whether the associated minimizers converge to the actual minimizers as the sample size increases. The positive answer to this question is crucial for the applicability of the proposed model aggregation scheme in real life concrete machine learning applications. This is particularly useful in contexts of online learning, such as the online surveillance of information networks, where information flows arrive from different sources (agents) and their validity must be frequently reassessed. In such settings, the respective weights for all nodes in the network need to be reallocated according to the performance of the provided estimates and the empirical evidence, therefore theoretical guarantees on their converging behaviour are important for the reliability of the aggregation mechanism.

\section{Consistency of the barycentric estimator under a variational setting}\label{sec-3}

In this section, we study the consistency properties of the proposed model aggregation scheme and the resulting barycentric estimator. Before embarking onto the technical aspects of this problem, let us briefly consider its qualitative aspects. As already mentioned above, the model aggregation problem reduces to a variational one as stated in \eqref{np}, whose data (i.e. the quantiles $q_j$ as well as the target quantile $q_{0}$) are random and depend on the sample used for infering them. Clearly, the estimates $q_{j,n}$ and $q_{0,n}$ based on the available sample are approximations of the true data of the problem $q_{j}$ and $q_{0}$, and it is expected that, as the sample size $n \to \infty$, the estimators $q_{j,n}, q_{0,n}$ converge (in the appropriate sense) to their true counterparts $q_{j}, q_{0}$. The consistency question can then be restated in the following way: Will the aggregate model corresponding to the empirical quantiles $q_{j,n}, q_{0,n}$ converge to the true aggregate model induced by the true quantiles $q_{j}, q_0$ in the limit, as the sample size $n \to \infty$? Using the formulation of barycenter calibration problem as a variational one (see Section \ref{sec-2}) we immediately see that the consistency problem as stated above, reduces to the problem of convergence of the minimizer induced by the empirical version of \eqref{np} to its actual (population-wise) minimizer. The convergence problem of minimizers of sequences of functionals to the minimizers of the true functional, is neither guaranteed in general nor is trivial. A positive answer to the aforementioned question requires the use of a particular notion of variational convergence called $\Gamma$-convergence, that was introduced in nonlinear analysis by Ennio de Giorgi (see e.g. \cite{dal2012introduction}). It is the aim of this section to show that the statistical consistency can be guaranteed under minimal assumptions, using the rigorous and general framework of $\Gamma$-convergence.

\subsection{Preliminaries on $\Gamma$-convergence}\label{sec-3.1}

First, we briefly provide some standard definitions concerning the $\Gamma$-convergence theory and the fundamental result concerning the convergence of minimizers. Let us denote by $( \Omega, \mathcal{F}, \mathbb{P})$ a probability space and consider a sequence of random functionals  $\{F_{n}\}_{n \in {\mathbb N}} := \{ F_n\}$, consisting of functionals  $F_n : \mathcal{X}\times \Omega \to \R \cup \{+\infty\}$ for $n \in \N$, where $\mathcal{X}$ is some metric space.  A definition of $\Gamma$-convergence for  the sequence of random functionals  $\{F_n \}$, in the spirit of \cite{thorpe2015convergence} and \cite{dunlop2020large}  is as follows.

\begin{definition}[$\Gamma$-convergence a.s.]\label{def-1}
	A sequence of random functionals $\{F_n\}$ $\Gamma$-converges almost surely (a.s.) to a functional $F:\mathcal{X}\times \Omega \to \R\cup\{+\infty\}$ if the following two conditions hold:
	\begin{itemize}
		\item[(LII)] For all $x \in \mathcal{X}$ and $x_n \to x$ a.s. it holds that 
		\begin{equation}\label{liminf}
			F(x,\omega) \leq \liminf_n F_n(x_n,\omega) \qquad \mathbb{P}-\mbox{a.s.}
		\end{equation}
		\item[(LSI)] For all $x\in\mathcal{X}$, there exists a (recovery) sequence $\widetilde{x}_n \to x$ a.s. for which holds that 
		\begin{equation}\label{limsup}
			F(x,\omega) \geq \limsup_n F_n(\widetilde{x}_n, \omega) \qquad \mathbb{P}-\mbox{a.s.}
		\end{equation} 	
	\end{itemize}
\end{definition}

A standard property that is employed when studying the convergence behaviour of minimization problems through the $\Gamma$-convergence (variational) framework is the coercivity property (and in particular equi-coercivity) of the  sequence of random functionals. The relevant definition follows.

\begin{definition}[Equi-coercivity a.s.]\label{equico-def}
	A sequence of random functionals $\{F_n\}$ is called equi-coercive almost surely if there exists a set $\Omega_0 \subset \Omega$ with $\mathbb{P}(\Omega_0) = 1$ such that, for every $\omega \in \Omega_0$ and every $\tau \in \R$, there exists a compact set $K \subset \mathcal{X}$ satisfying 
	$$ \{ x \in \mathcal{X} \, : \, F_n(x,\omega) \leq \tau \} \subset K$$ 
	for every $n \in \mathbb{N}$. 
\end{definition}

$\Gamma$-convergence for a sequence of functionals $\{F_n\}$  guarantees the convergence of the objective functional to be minimized, but on its own is not sufficient to guarantee the convergence of the sequence of minimizers $\{ x^*_n\}$  (where $x^*_n$ is the minimizer of $F_n$) to the minimizer $x^*$ of the limit functional $F$. This is exactly where the  equi-coercivity property comes into the game. The equi-coercivity property, rules out the possibility of pathological behaviour of the sequence of minimizers $\{x^*_n\}$ as $n \to \infty$, and guarantees the convergence of this sequence to the minimizer $x^*$, of the limit functional $F$ (the limit understood in the $\Gamma$-sense as in Definition  \ref{def-1}). Hence,  the combination of $\Gamma$-convergence and equi-coercivity yields convergence of minimum values and stability of minimizers; see e.g. \cite{dal2012introduction}. In Section \ref{sec-3.2}, we utilize the concept of $\Gamma$-convergence for the  treatment of the consistency question posed above. 
\color{black}

\subsection{$\Gamma$-convergence of the sequence of problems \eqref{np}}\label{sec-3.2}

Let us now denote by $(\Omega, \mathcal{F}, \mathbb{P})$ a probability space, related to the sampling procedure employed for generating  samples from the probability measures  $\mu_0, \mu_1, ..., \mu_J \in \mathcal{P}_{2}(\R)$.\footnote{We emphasize that the probability measure $\mathbb{P}$ (modelling the sampling procedure for obtaining the probability measures $\mu_i, \mu_0$)  is independent of the  probability measures $\mu_i, \mu_0$, themselves.} We consider the empirical version of the problem \eqref{np},  according to which samples of size $n$ are  used  to obtain estimates in terms of the empirical measures  $\mu_{j,n}(\omega), \mu_{0,n}(\omega)$ of the probability measures $\mu_{j}, \mu_{0}$ and the corresponding  empirical quantile functions in $L^2(0,1)$,  $q_{0,n}(\cdot,\omega)$, $q_{1,n}(\cdot,\omega), ..., q_{J,n}(\cdot,\omega)$ for  $\omega \in \Omega$. The dependence on $\omega$ indicates that the empirical measures, and the corresponding quantile functions, are random variables depending upon the sampling procedure. The subscript $n$ indicates the dependence of these entities on the sample size $n$. Our standing assumption (deliberately stated loosely for the time being; see Proposition \ref{prop-1}) is that (in the appropriate sense to be clarified shortly) $\mu_{j,n} \to \mu_j$, $\mu_{0, n} \to \mu_{0}$, ${\mathbb P}$-a.s.  in the large sample limit $n \to \infty$ (as well as for the corresponding quantile functions).

Based on the estimators for probability measures and in particular of the quantiles, $q_{j,n}, q_{0,n}$ for samples of size $n$, we  consider the empirical version of problem \eqref{np}. This is determined in terms of  the sequence $\{F_n\}$ of random functionals $F_{n} :   \R^{J} \times \Omega \to \R$ defined  for each $n$ as 
\begin{equation}\label{Jn}
	F_n(\w, \omega) = \left\{ 
	\begin{array}{lr} \frac{1}{2}\langle \w, {\bf G}_{n}(\omega) \, \w \rangle - \langle {\bf g}_{n}(\omega), \w \rangle + g_{0,n}(\omega),  & \,\,\, \w \in \SJ \\
		+\infty, &\,\,\, \w \notin \SJ
	\end{array}
	\right.
\end{equation}
where the random variables  ${\bf G}_{n} : \Omega \to  \R^{J\times J}_{+}$, ${\bf g}_{n} : \Omega \to \R^J$, $g_{0,n} : \Omega \to  \R$  are determined elementwise and pointwise in $\omega \in \Omega$ as
\begin{eqnarray*}
	({\bf G}_{n})_{i,j}(\omega) &:=& \langle q_{i,n}(\omega), q_{j,n}(\omega) \rangle = \int_0^1 q_{i,n}(s,\omega) \, q_{j,n}(s,\omega) ds, \,\,\, i,j \in \{1,2,...,J\}\\
	{\bf g}_{n,j}(\omega) &:=& \langle q_{j,n}(\omega), q_{0,n}(\omega) \rangle = \int_0^1 q_{j,n}(s,\omega) \, q_{0,n}(s,\omega) ds, \,\,\, j\in\{1,2,...,J\}\\
	g_{0,n}(\omega) &:=& \frac{1}{2} \langle q_{0,n}(\omega), q_{0,n}(\omega) \rangle = \frac{1}{2}\| q_{0,n}(\omega) \|_2^2 = \frac{1}{2}\int_0^1 q_{0,n}(s,\omega)^2 \, ds.
\end{eqnarray*}
According to our standing assumption, the sequences of random variables $\{{\bf G}_{n} \}, \{ {\bf g}_{n} \}, \{g_{0,n}\}$ converge ${\mathbb P}$-a.s. to the deterministic limits ${\bf G}$, ${\bf g}$, $g_0$, defined elementwise as 
$( {\bf G} )_{i,j} := \langle q_i, q_j \rangle = \int_0^1 q_i(s) q_{j}(s) ds$,  for $i, j \in \{ 1,2,...,J \}$, ${\bf g}_{i} := \langle q_i, q_0 \rangle$ for $i\in\{1,2,...,J\}$ and $g_0 := \frac{1}{2}\| q_0 \|_2^2$. Using the limits ${\bf G}$, ${\bf g}$, $g_0$,  we define the corresponding deterministic functional
\begin{equation}\label{Jlim}
	F(\w) = \left\{ 
	\begin{array}{lr} \frac{1}{2}\langle \w, {\bf G} \,\w \rangle - \langle {\bf g}, \w \rangle + g_{0}, & \w \in \SJ\\
		+\infty, & \w \notin \SJ. 
	\end{array} \right.
\end{equation}
This functional can be interpreted as the population objective, as opposed to the elements of the sequence $\{F_n\}$ that can be interpreted as the empirical objectives.  

Employing the abstract framework of Section \ref{sec-3.1}, for the choice ${\cal X}=\R^J$, our aim is to show that the sequence of random functionals $\{ F_n\}$ defined in \eqref{Jn}, in terms of the empirical quantiles involved, $\Gamma$-converges to the deterministic functional $F$, defined in \eqref{Jlim}, so that by guaranteeing also the equi-coercivity property for the sequence $\{ F_n\}$, we can show the convergence of the sequence $\{ \w_n\}$ of minimizers of the functionals in $\{F_n\}$ to the minimizer $\w$ of the limit functional $F$. This result, allows us to establishes the consistency  of the corresponding average models, with respect to the sampling procedure in the large sample limit ($n \to \infty$). This plan is established in the remainder of this section. In the interest of decluttering notation we omit the explicit dependence on $\omega \in \Omega$ for the elements of the sequences  $\{{\bf G}_{n} \}, \{ {\bf g}_{n} \}, \{g_{0,n}\}$.

The next proposition sets on a rigorous basis our standing assumption (stated in a rather loose way earlier on) of convergence of the empirical measures and other related quantities.
\color{black}

\begin{proposition}\label{prop-1}
	Let $\mu_{n,i}, \mu_i \in \mathcal{P}_2(\R)$ for $i=0,1,2,...,J$. Then, the random coefficients ${\bf G}_{n}:\Omega \to \R^{J \times J}$, ${\bf g}_n: \Omega \to \R^{J+1}$ and $g_{0,n}: \Omega \to \R$ defined in \eqref{Jn}, converge almost surely (elementwise) to the deterministic limits ${\bf G}\in \R^{J\times J}$, ${\bf g}\in\R^{J+1}$ and $g_0$ defined in \eqref{Jlim}.
\end{proposition}

\begin{proof}[Proof of Proposition \ref{prop-1}]	
	For the convergence of the random coefficients, it suffices to show that
	$$ \langle q_{i,n}, q_{j,n} \rangle \to \langle q_i, q_j \rangle \,\,\,\,  \mathbb{P}-\mbox{a.s.}$$
	for any $i,j \in \{0,1,...,J\}$. For almost every $\omega \in \Omega$ we work with the differences
	\begin{eqnarray*}
		|\langle q_{i,n}, q_{j,n}\rangle -  \langle q_{i}, q_{j}\rangle | 
		&=& |\langle q_{i,n} - q_i, q_{j,n}\rangle + \langle q_{j,n} - q_j, q_i \rangle| \\
		&\leq&  |\langle q_{i,n} - q_i, q_{j,n}\rangle| + |\langle q_{j,n} - q_j, q_i \rangle| \\
		&\leq& \|q_{i,n} - q_i\|_2 \| q_{j,n}\|_2 + \| q_{j,n} - q_j\|_2 \| q_i \|_2
	\end{eqnarray*}
	where by Lemma \ref{lem-1} in Appendix \ref{app-A} we obtain the convergence as $n\to\infty$. If we define $\Omega_i =\{\omega \, : \, \|q_{i,n}(\omega) - q_i\|_2 \to 0\}$ for $i=0,1,...,J$ where $\mathbb{P}(\Omega_i) = 1$ for all $i$, it holds that $\mathbb{P}(\bigcap_{i=0}^J \Omega_i) = 1$. Therefore, the convergence result holds $\mathbb{P}$-a.s. 
\end{proof}

Proposition \ref{prop-1} establishes the almost sure convergence of the random coefficients defining the empirical objective function $F_n$ to their population counterparts. Since both $F_n$ and $F$ are quadratic functions on the compact set $\SJ$, this coefficient-wise convergence implies uniform convergence of $F_n$ to $F$ on $\SJ$ almost surely. As shown below, this uniform convergence is sufficient to verify the $\Gamma$-convergence conditions of Definition \ref{def-1} in the present finite-dimensional and compact setting.

\begin{theorem}\label{th-1}
	Under the assumptions of Proposition \ref{prop-1}, it holds that:
	$$ F_n \xrightarrow{\Gamma} F \,\,\,\, \mbox{on}\,\, (\SJ, \|\cdot\|) \quad \mathbb{P}-\mbox{a.s.} $$
\end{theorem}

\begin{proof} 
	For any $\w \in \SJ$  we have ${\mathbb P}$-a.s. that
	\begin{eqnarray*}
		| F_n(\w) - F(\w) | &=& \left| \frac{1}{2} \langle \w, ( {\bf G}_{n} - {\bf G}) \w \rangle - \langle {\bf g}_n - {\bf g}, \w \rangle + (g_{n,0} - g_0) \right| \\
		&\leq& | \langle \w, ( {\bf G}_{n} - {\bf G}) \w \rangle | + | \langle {\bf g}_n - {\bf g}, \w \rangle| + |g_{n,0} - g_0| \\
		& \leq &\| ( {\bf G}_n - {\bf G}) \w \|_2 \,  \| \w \|_2  + \|  {\bf g}_n - {\bf g} \|_2 \| \, \w \|_2 + |g_{n,0} - g_0| \\
		&\leq & \| {\bf G}_n - {\bf G} \|_{*,2} \, \| \w \|_2^2 +  \|  {\bf g}_n - {\bf g} \|_2 \| \, \w \|_2 + |g_{n,0} - g_0|
	\end{eqnarray*}
	where the last inequality for the first term is obtained since if the spectral norm is used for the matrix ${\bf G}_{n} - {\bf G}$, by the relevant matrix norm definition $\| {\bf G}_{n} - {\bf G} \|_{*,2} := \sup_{ {\bf u } \ne 0} \frac{ \| ({\bf G}_{n} - {\bf G}) {\bf u}\|_2 }{ \| {\bf u} \|_2 }$ is implied that $\| ({\bf G}_{n} - {\bf G}) {\bf u}\|_2 \leq \| {\bf G}_{n} - {\bf G} \|_{*,2} \, \| {\bf u} \|_2$. Moreover, since $\|\w\|_2 \leq 1$ we get the upper bound 
	$$ 	| F_n(\w) - F(\w) | \leq \| {\bf G}_{n} - {\bf G} \|_{*,2}  + \| {\bf g}_n - {\bf g} \|_2  + | g_{n,0} - g_0 | $$ which is independent of $\w$. Therefore, taking supremum over $\SJ$ on  both sides and combining the result with Proposition \ref{prop-1} we get that
	\begin{equation}\label{UCR}
		\sup_{\w \in \SJ} | F_n(\w) - F(\w)| \to 0 \quad \mathbb{P}-\mbox{a.s.}.
	\end{equation}
	We will show that $\{F_n\}$ $\Gamma$-converges to $F$ by direct application of  Definition \ref{def-1}, employing the uniform convergence result established in \eqref{UCR}. 
	
	We first establish the $\Gamma-\liminf$ inequality (LII). Let us consider any $\{\w_n\} \subset \SJ$ such that $\w_n \to \w$. By the uniform convergence of $\{F_n\}$  to $F$ (see \eqref{UCR}) we have that for any fixed $\varepsilon > 0$, there exists $N \in \N$ such that for $n \geq N$  and any $\w' \in \SJ$, it holds that 
	$$F(\w') -\varepsilon < F_n(\w') < F(\w') + \varepsilon.$$ 
	Choosing $\w' = \w_n$ and taking liminf we get
	\begin{eqnarray*}
		\liminf_n F_n(\w_n) \geq \liminf_n (F(\w_n) - \varepsilon) = \liminf_n F(\w_n) - \varepsilon. 
	\end{eqnarray*}
	Since the above holds for any $\varepsilon > 0$ and by  the continuity of $F$ (it is a quadratic function  defined on the  compact set $\SJ$), we obtain
	\begin{equation}\label{eq-003}
		\liminf_n F_n(\w_n) \geq F(\w) \quad \mathbb{P}-\mbox{a.s.}
	\end{equation} 
	
	We next establish the $\Gamma-\limsup$ inequality (LSI) of Definition \ref{def-1}.  To this end we will show that, for any $\w \in \SJ$, the constant sequence $\{\widetilde{\w}_{n} \}$ defined by  $\widetilde{\w}_n := \w \in \SJ$ for any $n$ is a recovery sequence. By uniform convergence  (see \eqref{UCR}) we have that $F_n(\w) \to F(\w)$, i.e.
	$$ \limsup_n F_n(\widetilde{\w}_n) = \lim_n F_n(\w) = F(\w) $$
	and thus
	\begin{equation}\label{eq-004}
		F(\w) \geq \limsup_n F_n(\widetilde{\w}_n) \quad \mathbb{P}-\mbox{a.s.}
	\end{equation}
	Combining \eqref{eq-003} and \eqref{eq-004} we get that 
	$$ F_n \xrightarrow{\Gamma} F \quad \mathbb{P}-\mbox{a.s.} $$
\end{proof}
Having established that the sequence of empirical objective functions $\{F_n\}$ $\Gamma$-converges almost surely to the objective $F$ on the compact set $\SJ$, we now turn our attention to the corresponding minimization problems. Note that although the relevant minimization problems with respect to $\w$ stated in \eqref{np} (both the empirical and the limit problem) are quadratic, the uniqueness of their solutions is not guaranteed unless the matrices ${\bf G_n}, {\bf G}$ are strictly positive definite (in general they could also be positive semi-definite). Therefore, for the case of non-strictly convex functions $F_n, F$ the relevant minimization problem might have infinite minimizers. Our next result shows that the considered variational convergence translates into consistency of the minimization problems involved. In fact, for the general case (non-unique solution) we show that all empirical minimizers converge to limits that belong to the set of minimizers of the limit problem, while for the case of unique solutions (strictly convex objectives) we show that the (unique) solution obtained from the empirical barycenter calibration problem converges to the unique solution of the actual barycenter calibration problem. 
\begin{theorem}\label{th-2}
	Let $\w_n^{*} \in \arg\min_{\SJ} F_n$. Then, every limit point $\widehat{\w} \in \SJ$ of $\{\w^*_n\} \subset \SJ$ satisfies  
	$$\widehat{\w} \in \arg\min_{\w \in \SJ} F(\w) \quad \mathbb{P}-\mbox{a.s.}$$ 
	Moreover, if $F$ admits a unique minimizer $\w^* \in \SJ$, then  
	$$ \w_n^{*} \to \w^{*} \quad \mathbb{P}-\mbox{a.s.}$$
\end{theorem}

\begin{proof}
	Since $F_n$ are quadratic functions on the compact set $\SJ$ for any $n \in \N$, it is immediate that their minimizers exist $\mathbb{P}$-a.s. Let us denote
	$$ m := \inf_{w \in \SJ} F(\w) = F(\w^*), \qquad m_n := \inf_{\w \in \SJ} F_{n} = F_n(\w_n^*). $$
	By Theorem \ref{th-1} the random sequence $\{F_n\}$  (as in \eqref{Jn} ), $\Gamma$-converges $\mathbb{P}$-a.s. to the deterministic limit $F$  (as in \eqref{Jlim}) in $\SJ$. Moreover, for any $\tau \in \R$ it holds that 
	$$ B_{n,\tau} := \{ \w \in \SJ \, : \, F_n(\w, \omega) \leq \tau, \quad \mathbb{P}-\mbox{a.s.} \} \subset \SJ, \quad \forall n. $$
	Hence  it holds that $\bigcup_n B_{n,\tau} \subset \SJ$, and since $\SJ$ is compact, it holds that for every $\tau \in \R$, $\bigcup_n B_{n,\tau}$ is relatively compact $\mathbb{P}$-a.s. As a result, $\{F_n\}$ is $\mathbb{P}$-a.s. equi-coercive (please see Definition \ref{equico-def}). 
	
	We proceed with the proof in 4 steps.\\
	\noindent \emph{Step 1}: Let $\w^*_n \in \arg\min_{\SJ} F_n$ and let $\{n_k\}$ be such that $m_{n_k} \to \liminf_{n\to \infty} m_n$. By compactness of $\SJ$ there exists a further subsequence $\{\w^*_{n_{k_{\ell}}}\}$ such that  
	$$\w^*_{n_{k_{\ell}}} \to \widehat{\w} \in \SJ.$$
	Since $\{F_n\}$ $\Gamma$-converges a.s. to $F$ (by Theorem \ref{th-1}) we get by the $\Gamma$-liminf inequality that
	$$ F(\widehat{\w}) \leq \liminf_{\ell \to \infty} F_{n_{k_{\ell}}}(\w^*_{n_{k_{\ell}}}) = \liminf_{\ell \to \infty} m_{n_{k_{\ell}}} =  \liminf_{n \to \infty} m_{n}.$$
	Since $m \leq F(\widehat{\w})$ we get that 
	\begin{equation}\label{eq-001}
		m \leq \liminf_{n \to \infty} m_n.
	\end{equation}
	
	\noindent \emph{Step 2}: Let $\w^* \in \arg\min_{\SJ} F$. By the  $\Gamma$-$\limsup$ inequality (LSI) there exists a recovery sequence such that $\widetilde{\w}_n \to \w^*$ in $\SJ$ such that 
	$$ m = F(\w^*) \geq \limsup_{n \to \infty} F_n(\widetilde{\w}_n). $$
	However, since for every $n \in \N$ we have that $ m_n = \min_{\SJ} F_n \leq F_n(\widetilde{\w}_n)$ we get that
	\begin{equation}\label{eq-002}
		\limsup_{n \to \infty} m_n \leq \limsup_{n \to \infty} F_n( \widetilde{\w}_n ) \leq m.
	\end{equation}
	Combining \eqref{eq-001} and \eqref{eq-002} we get
	$$ m \leq \liminf_n m_n \leq \limsup_n m_n \leq m $$
	which implies that $m_n \to m$ (convergence of the minima).\\
	
	\noindent \emph{Step 3}: For a converging subsequence $\w^*_{n_k} \to \widehat{\w}$ we have that 
	$$ F(\widehat{\w}) \leq \liminf_k F_{n_k}(\w^*_{n_k}) = \liminf_k m_{n_k}.$$
	From Step 2 we have $m_{n_k} \to m$, and therefore $F(\widehat{\w}) \leq m$. But since by definition of $m$ we get the opposite inequality, we conclude that 
	$$ F(\widehat{\w}) = m $$
	and therefore, $\widehat{\w} \in \SJ$ is a minimizer of $F$.\\
	
	\noindent \emph{Step 4}: If $F$ admits a unique minimizer, then every $\{\w^*_{n_k}\} \subset \{\w^*_n\} \subset \SJ$ has a further subsequence $\w^*_{n_{k_\ell}} \to \w^*$. Then, necessarily we get $\w^*_n \to \w^*$. 
\end{proof}

\subsection{Statistical consistency of the barycentric estimator}\label{sec-3.3}

The results of Section \ref{sec-3.2} establish the consistency of the barycentric ensemble at the level of the weighting vector, showing that minimizers of the empirical objective functions converge to the minimizers of the population objective. We now translate these variational consistency results into consistency statements for the associated Wasserstein barycenters, at the level of the resulting barycentric estimator. We first establish uniform convergence for the  sequence of mappings $\w \mapsto \mu_{B, n}$ in $\SJ$, with respect to the weights of the empirical estimators (measures) to their population counterpart. 

\begin{proposition}\label{prop-2}
	Assume that for all $j=1,2,...,J$ it holds that $\mathcal{W}_2(\mu_{j,n}, \mu_j) \to 0$. Then,
	$$  \sup_{ \w \in \SJ} \mathcal{W}_2( \mu_{B,n}(\w), \mu_{B}(\w) ) \leq \max_{1\leq j \leq J} \mathcal{W}_2(\mu_{j,n}, \mu_j).$$
	In particular,
	$$ \sup_{ \w \in \SJ} \mathcal{W}_2( \mu_{B,n}(\w), \mu_{B}(\w) ) \to 0 \qquad \mathbb{P}-\mbox{a.s.}$$
\end{proposition}

\begin{proof}
	Recall that the estimators $\mu_{B,n}(\w), \mu_{B}(\w)$ can be represented in terms of their quantile functions
	$$ \qquad q_{B,n}(s;\w) = \sum_{j=1}^J w_j q_{j,n}(s), \qquad q_{B}(s;\w) = \sum_{j=1}^J w_j q_j(s), \qquad s\in(0,1),$$
	respectively. For any $\w \in \SJ$, fixed,  and almost every $\omega \in \Omega$ we have
	\begin{eqnarray*}
		\mathcal{W}_2( \mu_{B,n}(\w), \mu_{B}(\w) ) &=& \| q_{B,n}(\w) - q_{B}(\w) \|_2 \\ 
		&=& \| \sum_{j=1}^J w_j (q_{j,n} - q_j) \|_2 \\
		&\leq& \sum_{j=1}^J w_j \| q_{j,n} - q_j \|_2 \\
		&\leq&  \max_{1 \leq k \leq J} \| q_{k,n} - q_{k}\|_2 = \max_{1\leq j \leq J} \mathcal{W}_2(\mu_{j,n}, \mu_j)
	\end{eqnarray*}
	since $\w \in \SJ$ where $\sum_{j=1}^J w_j=1$, resulting to an upper bound which is independent of $\w$. Taking supremum over $\SJ$ in both sides, the upper bound still holds and converges to zero as $n \to \infty$ (by Lemma \ref{lem-1} in Appendix \ref{app-A}). 
\end{proof}

In Proposition \ref{prop-2} we established the uniform convergence of the individual empirical barycentric estimators to their population counterparts. We now combine this result with the variational convergence of the empirical weights studied in Section \ref{sec-3.2} to establish the almost sure consistency of the (induced) empirical barycenter estimator itself. 

\begin{theorem}\label{th-3}
	Let $\w^*_n \in \arg\min_{\SJ} F_n =: \mathbb{W}_{*,n}$, $\w^* \in \arg\min_{\SJ} F =: \mathbb{W}_*$ be arbitrary minimizers. Then, it holds that
	$$ \mathcal{W}_2( \mu^*_{B,n}, \mu^*_B ) \to 0 \qquad \mathbb{P}-\mbox{a.s.}$$
	where $\mu^*_{B,n} := \mu_{B,n}(\w^*_n)$, $\mu_B^* := \mu_{B}(\w^*)$.
\end{theorem}

\begin{proof}
	
	Employing Lemma \ref{lem-2} in Appendix \ref{app-A}, we have that for any element $\w^* \in \mathbb{W}_*$ and any $\w_n^* \in \mathbb{W}_{*,n}$ are induced unique barycentric estimators $\mu_B^*$ and $\mu_{B,n}^*$, respectively (the barycentric estimators are unique even if the sets $\mathbb{W}_*, \mathbb{W}_{*,n}$ are not singletons). Let any $\{ \mu_{B,n_k}(\w^*_{n_k})\}$. From compactness of $\SJ$, there exist subsequence $\w^*_{n_{k_{\ell}}} \to \widehat{\w} \in \SJ$. From Theorem \ref{th-2}, we have that $\widehat{\w} \in \arg\min_{\SJ} F = \mathbb{W}_*$. By Lemma \ref{lem-2} in Appendix \ref{app-A}, the induced population barycentric estimator is uniquely defined, so that $\mu_B^* = \mu_B(\widehat{\w})$ for any $\widehat{\w} \in \mathbb{W}_*$, i.e. for all the limits of the subsequences. Employing the triangle inequality we get
	\begin{eqnarray*}
		\mathcal{W}_2\big( \mu_{B,n_{k_{\ell}}}(\w^*_{n_{k_{\ell}}}), \mu_B^* \big) \, \leq \, \mathcal{W}_2\big( \mu_{B,n_{k_{\ell}}}(\w^*_{n_{k_{\ell}}}), \mu_B(\w^*_{n_{k_{\ell}}}) \big) \, + \, \mathcal{W}_2\big( \mu_B(\w^*_{n_{k_{\ell}}}), \mu_B^* \big)
	\end{eqnarray*}
	where the first component of the upper bound converges a.s. to 0 as $n\to\infty$ by Proposition \ref{prop-2} while the second component converges by continuity of the mapping $\w \mapsto \mu_B(\w)$. Since every subsequence $\{ \mu^*_{B, n_k} \}$ has converging subsequence $\{ \mu^*_{B, n_{k_{\ell}}} \}$ to $\mu_B^*$, the whole sequence converges, i.e. $ \mathcal{W}_2( \mu^*_{B,n}, \mu_B^* ) \to 0 \qquad \mathbb{P}-\mbox{a.s.}$
\end{proof}

The previous results show that the proposed barycentric procedure is asymptotically stable both in terms of the optimal weights and the induced probability distributions. In particular, variational convergence of the empirical optimization problems transfers to convergence of the associated barycentric estimators under the quadratic Wasserstein metric.

\section{Synthetic data experiments}\label{sec-4}

This section evaluates the finite-sample performance of the barycentric calibration estimator defined in \eqref{np} in a controlled setting where the target distribution is itself a Wasserstein barycenter of a finite model family. The main goal is to assess whether the estimated weights $\widehat \w$ can recover the true weights
$\w_{true}$, and how close the resulting barycenter $\widehat q$ is to (i) the empirical
quantile $q_{0,n}$ and (ii) the true quantile $q_{\mathrm{true}}$.

\subsection{Model families and the data-generating mechanism} 

For each configuration we fix: (i) a sample size $n$, (ii) a number of candidate models $J$, and (iii) a distributional family type. The candidate family consists of parametric models whose quantile functions are evaluated on a common grid. We consider the following libraries:
\begin{itemize}
	\item \textbf{Weibull:} all $J$ candidates are Weibull.
	\item \textbf{Gamma:} all $J$ candidates are Gamma.
	\item \textbf{Mixed:} the $J$ candidates contain fixed counts of Weibull and Gamma models (summing to $J$), with the order randomly permuted.
\end{itemize}
Within each replication, for each candidate model $j\in\{1,\dots,J\}$ we generate a scale parameter independently as
$$ a_j \sim \mathrm{Unif}(a_{\mathrm{lo}},a_{\mathrm{up}}),
\qquad a_{\mathrm{lo}}=0.7,\;\; a_{\mathrm{up}}=1.3.
$$
In addition, we consider three baseline shape levels
$$
s \in \{0.7,\;1.0,\;1.3\}.
$$
Conditional on the chosen baseline $s$, we randomize the candidate-specific shape via a multiplicative perturbation
$$ \eta_j \sim \mathrm{Unif}(\underline m_s,\overline m_s), \qquad \kappa_j = s\,\eta_j, $$
where
$$ (\underline m_{0.7},\overline m_{0.7})=(0.4,1.6), \qquad (\underline m_{1.0},\overline m_{1.0})=(0.7,1.3),\qquad (\underline m_{1.3},\overline m_{1.3})=(0.6,1.4). $$
The resulting quantity $\kappa_j$ is used as the shape parameter of the $j$-th candidate in the Weibull and Gamma cases. Denoting by $q_j$ the candidate quantile values, we compute $q_j$ using the generalized-gamma quantile routine from the \texttt{flexsurv} package in R, which provides a unified interface across families. For each configuration $(n,J)$ we generate a fixed true weight vector $\w_{true}$ on the unit simplex, by drawing positive weights, and normalizing them so that $\sum_{j=1}^J w_{j,true}=1$. This fixed $\w_{true}$ is held constant across replications within the same configuration so that weight recovery is meaningful. Within each replication, after generating the candidate quantiles $\{q_j\}_{j=1}^J$, we define the true barycenter quantile curve by
$$ q_{\mathrm{true}}(u) = \sum_{j=1}^J w_{j,true}\, q_j(u), \quad u \in (0,1).$$ 
Thus, the target distribution is exactly a Wasserstein barycenter of the candidate library with weights $\w_{\mathrm{true}}$. To generate observations from this target, we use an inverse transform sampling process, i.e. 
$$U_1,\dots,U_n \stackrel{\text{i.i.d.}}{\sim}\mathrm{Unif}(0,1), \qquad 
X_{i} = q_{\mathrm{true}}(U_{i}),\qquad i=1,\dots,n.$$
From the generated sample $X_1,\dots,X_n$ we compute the empirical quantile $q_{0,n}$.

\subsection{Weight estimation via barycentric calibration and performance criteria}\label{subsec:estimation}

Let $q_B(u_\ell;w)=\sum_{j=1}^J w_j q_j(u_\ell)$ denote the barycenter quantile induced by weights $\w\in\Delta_J$ where $\{u_{\ell}\}_{\ell=1}^K$ denotes a grid of size $K$ in $(0,1)$. For distributions in the real line, the squared Wasserstein distance equals the $L^2$ distance between quantile functions. We therefore estimate weights by solving the discrete analogue of \eqref{np}:
\begin{equation}\label{eq:discrete_calib}
	\widehat \w \in \arg\min_{\w \in \Delta_J}\;
	\frac{1}{2}\cdot \frac{1}{K}\sum_{\ell=1}^K \Big(q_B(u_\ell; \w) - q_{0,n}(u_\ell)\Big)^2.
\end{equation}
We then define the fitted barycenter quantile as
$$ \widehat q(u_\ell)=q_B(u_\ell;\widehat \w). $$
Then, on each replication the following performance indices are recorded:
\begin{itemize}
	\item  Empirical fit: 
	$$ W_{2,\mathrm{emp}} = \left(\frac{1}{K}\sum_{\ell=1}^K\big(\widehat q(u_\ell)-q_{0,n}(u_\ell)\big)^2\right)^{1/2}. $$
	
	\item $W_2$ distance from the actual model: 
	$$ W_{2,\mathrm{true}} =
	\left(\frac{1}{K}\sum_{\ell=1}^K\big(\widehat q(u_\ell)-q_{\mathrm{true}}(u_\ell)\big)^2\right)^{1/2}.$$
	\item  Weight-recovery errors ($L^1$, $L^2$ and $L^{\infty}$ norms): 
	$$ \|\widehat \w-\w_{\mathrm{true}}\|_1,\qquad
	\|\widehat \w-\w_{\mathrm{true}}\|_2,\qquad
	\|\widehat \w-\w_{\mathrm{true}}\|_\infty. $$
\end{itemize}

\subsection{Simulation results}\label{results}

In this subsection, we report the results illustrated in Tables \ref{tab-s1}, \ref{tab-s2} and \ref{tab-s3} in the Appendix. For each configuration (fixed $n$, $J$, model family and shape scenario), the corresponding performance statistics are calculated over a sample of 1000 replications. For each performance metric, the mean and standard error (in parentheses) values are given. Across all configurations, both $W_{2,\mathrm{emp}}$ and $W_{2,\mathrm{true}}$ decrease
systematically as $n$ increases. This monotone improvement reflects the variational consistency results of Section \ref{sec-3}: as the empirical quantile curve $q_{0,n}$ stabilizes, the calibrated barycenter $\widehat q$ approaches the barycentric target $q_{\mathrm{true}}$.

In addition, $W_{2,\mathrm{emp}}$ is typically smaller than $W_{2,\mathrm{true}}$, which is natural, since $W_{2,\mathrm{emp}}$ measures the fit to $q_{0,n}$ (which is a noisy approximation of $q_{\text{true}}$ at finite $n$), whereas $W_{2,\mathrm{true}}$
measures the discrepancy from the target $q_{\mathrm{true}}$. Importantly, the magnitude of the improvement with $n$ is substantial. For instance, in the Gamma case with $J=5$, $W_{2,\mathrm{true}}$ decreases from order $10^{-1}$ at $n=100$ to order $10^{-2}$ at $n=10000$ across all shape scenarios, with similarly stable behavior in the Mixed case. These patterns confirm that barycentric calibration provides a Wasserstein-consistent estimator even when the candidate family is heterogeneous.

A key empirical message is the separation between distributional accuracy and weight identification. Increasing the size $J$ has only a modest effect on $W_{2,\mathrm{emp}}$ and $W_{2,\mathrm{true}}$ (for moderate to large $n$), while it has a clearer effect on the weight-error norms. This is expected and does not contradict the methodological goal: in barycentric model averaging, different weight vectors can generate very similar (and sometimes nearly indistinguishable) barycenters, especially when the candidate quantile curves are highly correlated or redundant. Thus, exact recovery of $\w_{\mathrm{true}}$ is intrinsically more demanding than recovering
the barycentric target distribution. The aforementioned Tables indicate that, even when weight recovery becomes harder as $J$ grows, the fitted barycenter $\widehat q$ remains accurate in terms of the Wasserstein distance, which is the primary object of inference in our framework. On the other hand, the most challenging regime is the Weibull family with baseline shape $s=0.7$. In this setting, both $W_{2,{\text{emp}}}$ and $W_{2,{\text{true}}}$ remain comparatively large, even at $n=10000$, and the weight errors are also elevated. However, the method still exhibits the correct direction of improvement with $n$ (both $W_{2,\mathrm{emp}}$ and $W_{2,\mathrm{true}}$ decrease), and performance is markedly improved for the other shape levels ($s=1.0$ and $s=1.3$).

Overall, based on our simulations, we conclude that: (i) barycentric calibration recovers a barycenter close to the empirical quantile curve ($W_{2,{\text{emp}}}$ small for moderate/large $n$), and (ii) it yields a barycenter close to the true generating barycenter ($W_{2,{\text{true}}}$ decreasing with $n$), even under heterogeneous candidate model families (Mixed case).


\section{Maximum temperature distributions monitoring on a spatial domain}\label{sec-5}

In this Section, a real data application is perfomed employing the discussed aggregation framework for monitoring the distributions of maximum temperature over a spatial network of stations. The proposed barycentric estimator is employed to optimally aggregate the information provided by the observable nodes of the network (i.e. the ones referred to as training sites) to recover the actual maximum temperature distributions of the unobservable nodes (i.e. the ones referrd to as test sites). Two versions of the barycentric estimator are employed for performance comparison reasons: (a) the Wasserstein barycenter for which the weights remain constant and uniformly allocated (i.e. $w_j = 1/J$ for all $j=1,2,...,J$) referred to as the uncalibrated barycenter, and (b) the calibrated barycentric estimator which weights are determined by the solution of problem \eqref{np}. 

\begin{figure}[ht!]
	\centering
	\includegraphics[width=3in]{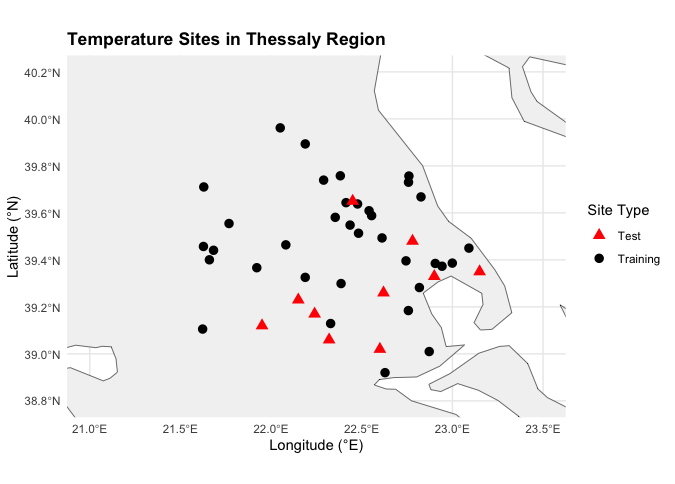}			
	\caption{The chosen temperature sites in the area of Thessaly (Greece). Black circles represent the training sites (known sources), while the red triangles represent the test sites (unkown sources, to be recovered).}\label{fig-1}
\end{figure}

The dataset used in this study are part of the CLIMADAT-GRid dataset \citep{varotsos_2025}, a gridded climate product for Greece that covers the years 1981-2019 at a spatial resolution of $1\,\text{km} \times 1\,\text{km}$. It is constructed from quality-controlled and homogenized daily measurements from a large network of meteorological stations (122 for temperature and 312 for precipitation). The original dataset provides the following variables: daily precipitation sums (pr, mm), daily mean air temperature (tg, $^\circ$C), daily maximum air temperature (tx, $^\circ$C), daily minimum air temperature (tn, $^\circ$C), surface altitude (orog, m), and a land-sea mask (mask, 1). In the present analysis, we focus on daily maximum temperature (tx) for the region of Thessaly (Figure \ref{fig-1}). For the purposes of the study, a subset of $J=35$ sites are assigned as training sites (listed with their spatial coordinates in Table \ref{training_sites} in the Appendix) and a subset of 10 sites are assigned as test sites (listed in Table \ref{tab:unknown_sites} in the Appendix with their spatial coordinates and a briefly description of the features in each loaction). The information batches from the test sites are provided only at certain time instants (when the weighing vectors are reallocated) and the main task of the aggregation scheme is to optimally aggregate the information provided by the rest $J=35$ sites (which are considered as on-line data providers) to approximate the true distribution of each one of the test sites. Well–distributed training sites ensure that their barycenter captures a broad range of local climatological behaviours, which is critical for accurate spatial transfer to the test stations. In order to evaluate the performance of the method, the aggregation task is repeated for successive years. For each site and year we represented the empirical distribution of daily maximum temperature by its quantile function evaluated on a fine grid of quantile levels. A rolling window of three years has been used for the calibration task. Prediction performance has been assessed by the quadratic Wasserstein distance. 

\begin{figure}[ht!]
	\centering
	\includegraphics[width=5in]{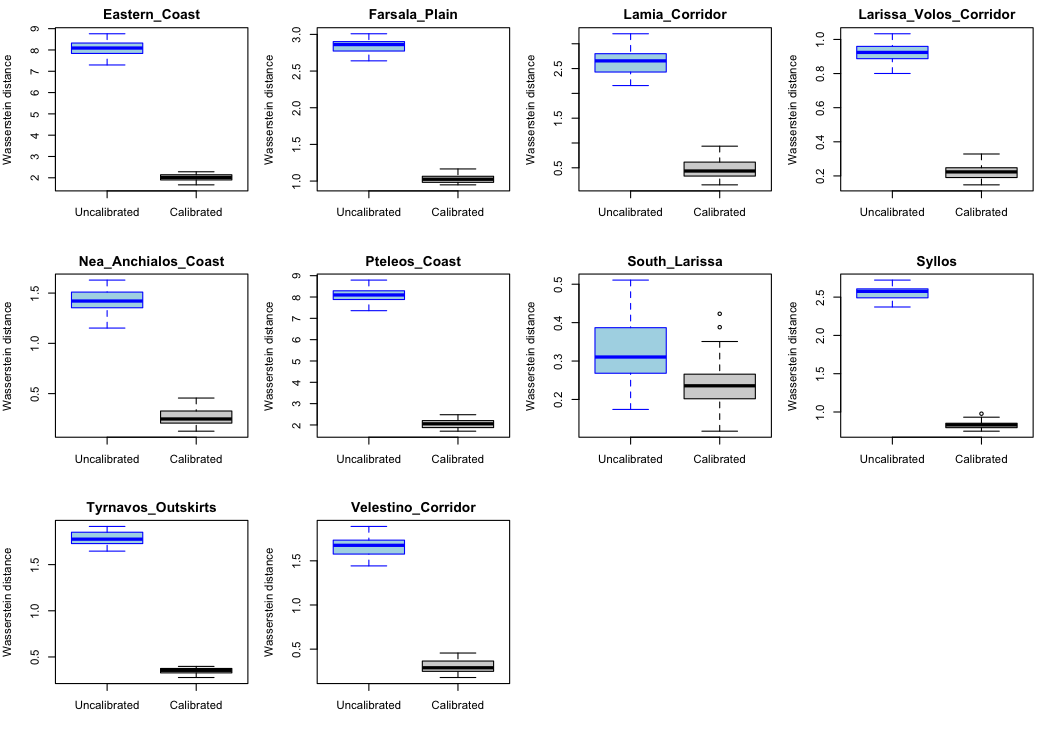} 
	\caption{Wasserstein distance from the true model boxplots for the uniformly weighted barycenter (gray boxes) and the calibrated barycenter (blue boxes) per site for the prediction period (1984 - 2019)}\label{W2dist}
\end{figure}

\begin{figure}[ht!]
	\centering
	\includegraphics[width=5in]{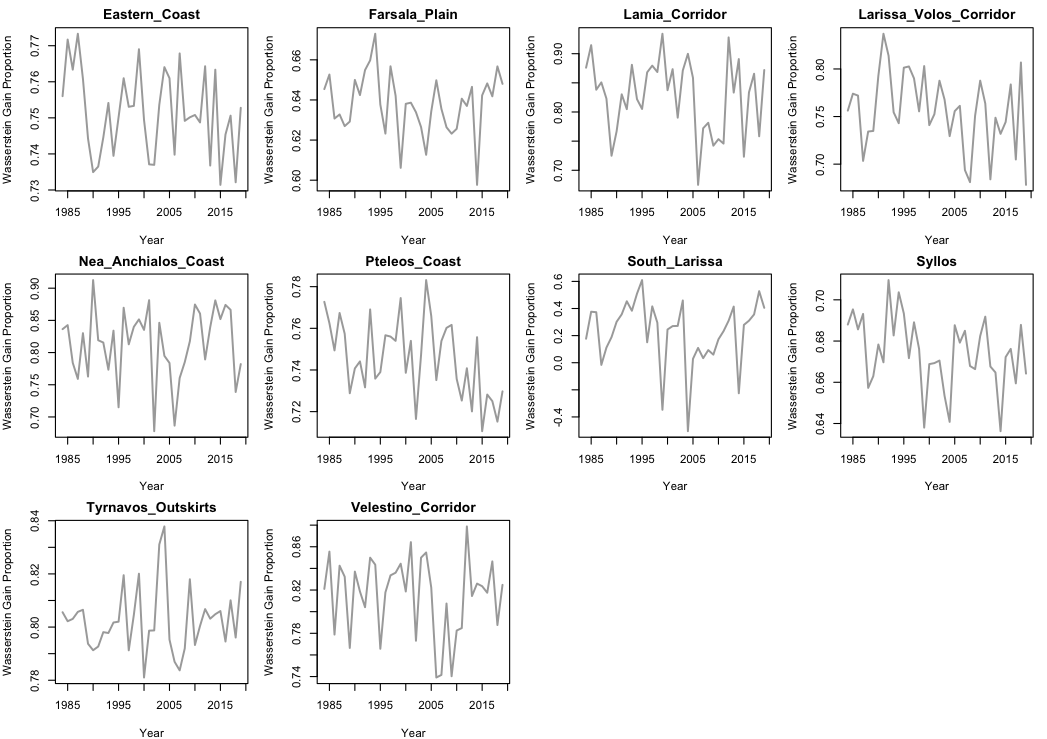} 
	\caption{Wasserstein gain proportion plots per site for the prediction period (1984 - 2019)}\label{W2gain}
\end{figure}

In Figure \ref{W2dist}, are illustrated for each test site the observed Wasserstein discrepancies of the calibrated and uncalibrated barycentric estimators from the observed distributions. The superiority of the calibrated barycenters (gray boxplots) comparing to the uniformly weighted (blue boxplots) is clear to all cases. In fact, in the nine of the ten cases the boxplots of the calibrated barycenters lie much lower than the uniformly weighted ones, indicating much lower approximation error. Even in the one case (South Larissa region) that the difference is smaller, the calibrated model has in general smaller range in the approximation error and the whole boxplot lies lower. 

The superiority on the approximation performance of the calibrated barycenter becomes much clearer in Figure \ref{W2gain}. In this figure is illustrated per prediction year the gain proportion on employing the calibrated version of the barycentric estimator instead of the uniformly weighted counterpart and measured by the performance metric
$$ \mathrm{Wasserstein\, Gain \, Proportion}_y = \left( 1 - \frac{ \mathrm{Dist}^{*}_{y} }{  \mathrm{Dist}^{\mathrm{unif}}_{y} } \right) $$
where $y$ denotes the year, $\mathrm{Dist}^{*}_{y}$ denotes the Wasserstein discrepancy of the observed distribution at $y$ from the calibrated barycentric model and $\mathrm{Dist}^{\mathrm{unif}}_{y}$ denotes the discrepancy from the uniformly weighted counterpart. It is clear that in the majority of the cases, the calibrated aggregate model reduces the distance from the actual distribution at least 60\% and 70-75\% on average of the years (we exclude the South Larissa region where the improvement is not so impressive, but still it is about 20-25\% on average). Moreover, in Appendix in Figure \ref{VaR_0950} and \ref{VaR_0990} is illustrated the approximation performance of the estimators with respect to the upper tail values, and in particular the 95\% and 99\% Value at Risk (VaR). The approximation behaviour of the calibrated barycentric estimator remains superior even in the upper tail values of the distribution. Overall, the calibrated barycentric estimator distributional error displays much less magnitude than the uniformly weighted one's, indicating the importance of the barycenter's calibration task in its approximation performance.

\section{Conclusions}\label{sec-6}

In this work, we studied from a variational perspective the problem of model aggregation in the space of probability distributions through Wasserstein barycenters. Based on the notion of $\Gamma$-convergence, we provided consistency of the resulting barycentric estimator on a variational context which is the appropriate one when the involved estimators and its parameters both depend on the available empirical evidence and the sample size. Under the minimum regularity assumptions, we established consistency results for the related empirical minimization problems and the induced estimators. The qualitative properties of the proposed aggregation process and the performance of the proposed estimator were studied through two concrete applications. First, the consistency of the aggregation scheme and its approximation performance are studied in extensive and carefully designed synthetic data experiments, where the effects of the sample size and the cardinality of the set of available models, are properly assessed and found to be in line with the provided theoretical results. Finally, the method was successfully implemented for the approximation of maximum temperature distributions, in a spatial network with a broad range of climatological conditions. The effect of the optimal choice of the weighting vector in the performance of the employed barycentric estimator proved extremelly significant, since it completely outperformed the competiting benchmark model.

\bibliographystyle{chicago} 
\bibliography{bibliography}

\appendix

\newpage
\section{Supplementary material from Section \ref{sec-3}}\label{app-A}

\begin{lemma}\label{lem-1}
	Let $X_1, X_2, ..., X_n$ be a random sample with independent observations from $\mu \in \mathcal{P}_2(\R)$, and let $ \mu_n := \frac{1}{n}\sum_{i=1}^n \delta_{X_i}$
	be the induced empirical measure. Denote by $q_{n}$ and $q$ the quantile functions of $\mu_n$ and $\mu$, respectively. Then, 
	$$ \| q_{n} - q \|_2 \to 0 \qquad \mathbb{P}-a.s. $$
\end{lemma}

\begin{proof}[Proof of Lemma \ref{lem-1}] Let $F_{n}$ and $F$ denote the distribution functions of the probability measures $\mu_n$ and $\mu$, respectively. By the Glivenko-Cantelli theorem we get 
	$$\sup_{x\in\R} |F_{n}(x) - F(x)| \to 0 \qquad \mathbb{P}-\mbox{a.s.} \qquad \mbox{ as } n \to \infty $$ 
	which implies the almost surely weak convergence
	\begin{equation}\label{cond-1}
		\mu_{n} \to \mu \qquad \mathbb{P}-\mbox{a.s.}
	\end{equation} 
	Since $\mathbb E[X_1^2]<\infty$, the Strong Law of Large Numbers (SLLN) applied to
	$X_i^2$ yields
	\begin{equation}\label{cond-2}
		\int_{\mathbb R} x^2\,d\mu_n(x) = \frac1n\sum_{i=1}^n X_i^2 \longrightarrow \mathbb E[X_1^2] = \int_{\mathbb R} x^2\,d\mu(x) \quad \mathbb P\text{-a.s.}
	\end{equation}
	Combining \eqref{cond-1}, \eqref{cond-2} and Prop. 7.1.5 in \cite{ambrosio2008gradient} implies that 
	$$\mathcal{W}_2(\mu_{n}, \mu) \to 0 \qquad \mathbb{P}-\mbox{a.s.}$$
	In the real line, $\mathcal{W}_2(\mu_n, \mu) = \| q_{n} - q \|_2$, hence equivalently we get	
	\begin{equation}\label{cond-3}
		\| q_{n} - q \|_2 \to 0 \qquad \mathbb{P}-\mbox{a.s.}
	\end{equation} 
\end{proof}

\begin{lemma}\label{lem-2}
	Let $\mathbb{W}_* := \arg\inf_{\w \in \SJ} F(\w)$. For any $\w^*_0 \in \mathbb{W}_*$ it holds that 
	$$ \mathbb{W}_* = \big\{ \w^*_0 + {\bf h} \, : \, {\bf h} \in \ker({\bf G}) \big\} \cap \SJ $$
	where $\ker({\bf G}) = \{ {\bf h} \in \R^J \, : \, {\bf G} {\bf h} = {\bf 0} \ \}$. Moreover, it holds that $$\mu_B^* := \mu_B(\w^*_0) = \mu_B(\w^*), \quad \forall \w^* \in \mathbb{W}_*$$ 
	that is, the barycentric estimator is uniquely defined even when the optimal weights are not unique.
\end{lemma}

\begin{proof}
	We consider the general case where the matrix ${\bf G}$ is positive semidefinite, i.e. the function $F$ as determined in \eqref{Jlim} is convex (and not strictly convex) on $\w$. Let $\w_0^*, \w^* \in \mathbb{W}_*$ and therefore $F(\w_0^*) = F(\w^*) = \min_{\SJ} F(\w) =: m$. Define also ${\bf h} := \w^* - \w_0^*$ and 
	$$ \w_{\lambda} := \w_0^* + \lambda {\bf h} = (1-\lambda) \w_0^* + \lambda \w^* \in \SJ, \quad \lambda \in [0,1]. $$
	By convexity of $F$ it holds that
	$$ F( \w_{\lambda} ) \leq \lambda F(\w^*) + (1-\lambda) F(\w_0^*) = m. $$
	Moreover,  for any $\lambda \in [0,1]$ it holds that $F(\w_{\lambda}) \geq m$, and combining with the above we get that $F(\w_{\lambda}) = m$. Therefore any $\w_{\lambda}$ is also a minimizer of $F$. Then, calculating $F$ at any $\w_{\lambda}$, we get
	\begin{equation}\label{expansion}
		F(\w_{\lambda}) = F( \w_0^* + \lambda {\bf h} ) = F(  \lambda \w^* + (1-\lambda) \w_0^* ) = F(\w_0^*) + \lambda \langle {\bf G} \w_0^* - {\bf g}, {\bf h} \rangle + \frac{\lambda^2}{2} \, {\bf h}^T {\bf G} {\bf h}.
	\end{equation}	
	Since $F(\w_0^*) = F(\w_{\lambda}) = m$ for all $\lambda$, we necessarily have that 
	\begin{eqnarray*}
		&& \lambda \langle {\bf G} \w_0^* - {\bf g}, {\bf h} \rangle + \frac{\lambda^2}{2} \, {\bf h}^T {\bf G} {\bf h} = 0, \quad \forall \lambda \in [0,1] \iff \\
		&& \lambda \big( {\bf h}^T {\bf G} \w_0^* - {\bf g}^{T} {\bf h} + \frac{\lambda}{2} {\bf h}^T {\bf G} {\bf h} \big) = 0,  \quad \forall \lambda \in [0,1] \iff \\
		&& {\bf h}^T {\bf G} \w_0^* - {\bf g}^{T} {\bf h} + \frac{\lambda}{2} {\bf h}^T {\bf G} {\bf h} = 0, \quad \forall \lambda \in (0,1].
	\end{eqnarray*}
	Since the above identity holds for all $\lambda \in [0,1]$, the affine function $\lambda \mapsto {\bf h}^T {\bf G} \w_0^* - {\bf g}^{T} {\bf h} + \frac{\lambda}{2} {\bf h}^T {\bf G} {\bf h}$ must vanish identically. Hence
	$$ {\bf h}^T {\bf G} \w_0^* - {\bf g}^{T} {\bf h} = 0 \quad \mbox{ and } \quad {\bf h}^T {\bf G} {\bf h} = 0.$$
	Since ${\bf G}$ is positive semidefinite, if it holds that ${\bf h}^T {\bf G} {\bf h} = 0$ then ${\bf G}{\bf h} = {\bf 0}$ which leads to ${\bf h} = \w^* - \w_0^* \in \ker({\bf G})$. Since any element of $\mathbb{W}_*$ belongs also to $\SJ$, we get that
	\begin{equation}\label{left-inc}
		\mathbb{W}_* \subseteq \{ \w_0^* + {\bf h} \, : \, {\bf h } \in  \ker({\bf G} ) \} \cap \SJ.
	\end{equation}  
	
	\noindent Conversely, let $\w^* := \w_0^* + {\bf h} \in \{ \w_0^* + {\bf h} \, : \, {\bf h } \in  \ker({\bf G} ) \} \cap \SJ$. Therefore, it holds that ${\bf G} {\bf h} = {\bf 0}$ and since ${\bf G}$ is symmetric we have that
	$$ {\bf h}^T {\bf G} {\bf h} = 0, \quad {\bf h}^T {\bf G} \w_0^* = 0. $$
	We just need to show that the term ${\bf h}^T {\bf g}$ in \eqref{expansion} vanishes. Expanding the term we have that 
	$$ {\bf h}^T {\bf g} = \big\langle \sum_{j=1}^J h_j q_j, q_0 \big\rangle = \langle q_B({\bf h}), q_0 \rangle.$$
	But, from the condition ${\bf h}^T {\bf G} {\bf h} = 0$ we get that 
	$$ {\bf h}^T {\bf G} {\bf h}  = \bigg\| \sum_{j=1}^J h_j q_j \bigg\|_2^2 = 0 \quad \Rightarrow \quad q_B({\bf h}) = \sum_{j=1}^J h_j q_j(s) = 0, \,\, \mbox{a.e. on }(0,1) \quad \Rightarrow \quad {\bf h}^T {\bf g} = 0.$$
	Therefore, we get that $F(\w_0^*) = F(\w^*) = F(\w_0^* + {\bf h}) = m$ and therefore each $\w^*$ is also a minimizer. Since $\w^* \in \SJ$ we have that $\w^* \in \mathbb{W}_*$, i.e. it also holds that 
	\begin{equation}\label{right-inc}
		\mathbb{W}_* \supseteq \{ \w_0^* + {\bf h} \, : \, {\bf h } \in  \ker({\bf G} ) \} \cap \SJ.
	\end{equation}	
	By relations \eqref{left-inc} and \eqref{right-inc} we get that $	\mathbb{W}_* = \{ \w_0^* + {\bf h} \, : \, {\bf h } \in  \ker({\bf G} ) \} \cap \SJ.$ Clearly, if $F$ is strictly convex (i.e. ${\bf G}$ positive definite which implies $\ker({\bf G}) = {\bf 0}$) the set $\mathbb{W}_*$ is a singleton.\\ 
	
	\noindent Moreover, if $\w^* = \w_0^* + {\bf h}$ with ${\bf h} \in \ker( {\bf G} )$ we have that
	\begin{eqnarray*}
		\mathcal{W}_2^2( \mu_B(\w^*), \mu_B(\w_0^*) ) = \| q_B(\w^*) - q_B(\w_0^*) \|_2^2 = \big\| \sum_{j=1}^J h_j \, q_j \big\|_2^2 = \| q_B({\bf h})\|_2^2 = 0  
	\end{eqnarray*}
	and therefore $\mu_B(\w^*) = \mu_B(\w_0^*)$ for any $\w^* \in \mathbb{W}_*$.	
\end{proof}

\newpage
\section{Supplementary material from Section \ref{sec-4}}\label{app-B}

\begin{table}[ht!]
	\centering
	\caption{Weight-recovery summaries for Weibull model simulation study.}\label{tab-s1}
	\centering
	\fontsize{7}{11}\selectfont
	\begin{tabular}[t]{rrlllllll}
		\toprule
		n & J & distribution & shape & $W_{2,\mathrm{emp}}$ & $W_{2,\mathrm{true}}$ & $L^1$ & $L^2$ & $L^\infty$\\
		\midrule
		100 & 5 & Weibull & 0.7000 & 1.4569 (0.0611) & 1.6910 (0.0719) & 1.0197 (0.0105) & 0.5446 (0.0061) & 0.4283 (0.0060)\\
		500 & 5 & Weibull & 0.7000 & 0.7329 (0.0282) & 0.8431 (0.0353) & 0.7310 (0.0108) & 0.3892 (0.0060) & 0.2963 (0.0054)\\
		1000 & 5 & Weibull & 0.7000 & 0.5330 (0.0200) & 0.7672 (0.0346) & 0.6324 (0.0097) & 0.3353 (0.0052) & 0.2518 (0.0045)\\
		10000 & 5 & Weibull & 0.7000 & 0.2867 (0.0105) & 0.6200 (0.0252) & 0.5800 (0.0084) & 0.3055 (0.0045) & 0.2262 (0.0039)\\
		\addlinespace
		100 & 5 & Weibull & 1.0000 & 0.1590 (0.0020) & 0.1398 (0.0028) & 0.8352 (0.0106) & 0.4373 (0.0058) & 0.3271 (0.0052)\\
		500 & 5 & Weibull & 1.0000 & 0.0736 (0.0009) & 0.0729 (0.0014) & 0.4967 (0.0075) & 0.2591 (0.0038) & 0.1854 (0.0030)\\
		1000 & 5 & Weibull & 1.0000 & 0.0534 (0.0006) & 0.0571 (0.0011) & 0.4077 (0.0056) & 0.2128 (0.0029) & 0.1517 (0.0023)\\
		10000 & 5 & Weibull & 1.0000 & 0.0275 (0.0003) & 0.0310 (0.0005) & 0.2832 (0.0031) & 0.1465 (0.0016) & 0.1031 (0.0012)\\
		\addlinespace
		100 & 5 & Weibull & 1.3000 & 0.1014 (0.0013) & 0.0943 (0.0019) & 0.6978 (0.0097) & 0.3653 (0.0051) & 0.2674 (0.0043)\\
		500 & 5 & Weibull & 1.3000 & 0.0471 (0.0006) & 0.0474 (0.0009) & 0.4064 (0.0062) & 0.2128 (0.0031) & 0.1512 (0.0023)\\
		1000 & 5 & Weibull & 1.3000 & 0.0342 (0.0004) & 0.0364 (0.0007) & 0.3380 (0.0046) & 0.1761 (0.0023) & 0.1250 (0.0018)\\
		10000 & 5 & Weibull & 1.3000 & 0.0168 (0.0002) & 0.0177 (0.0003) & 0.2359 (0.0027) & 0.1216 (0.0013) & 0.0851 (0.0009)\\
		\midrule
		100 & 10 & Weibull & 0.7000 & 1.3625 (0.0461) & 1.6369 (0.0545) & 1.1539 (0.0114) & 0.4886 (0.0060) & 0.3588 (0.0063)\\
		500 & 10 & Weibull & 0.7000 & 0.7038 (0.0212) & 0.9487 (0.0332) & 0.8607 (0.0105) & 0.3430 (0.0047) & 0.2269 (0.0042)\\
		1000 & 10 & Weibull & 0.7000 & 0.4693 (0.0137) & 0.7464 (0.0257) & 0.7518 (0.0087) & 0.2931 (0.0037) & 0.1853 (0.0032)\\
		10000 & 10 & Weibull & 0.7000 & 0.2540 (0.0071) & 0.6427 (0.0188) & 0.7036 (0.0072) & 0.2695 (0.0029) & 0.1647 (0.0022)\\
		\addlinespace
		100 & 10 & Weibull & 1.0000 & 0.1528 (0.0018) & 0.1533 (0.0028) & 0.9147 (0.0097) & 0.3646 (0.0043) & 0.2374 (0.0040)\\
		500 & 10 & Weibull & 1.0000 & 0.0726 (0.0008) & 0.0762 (0.0014) & 0.6166 (0.0054) & 0.2355 (0.0021) & 0.1387 (0.0015)\\
		1000 & 10 & Weibull & 1.0000 & 0.0520 (0.0005) & 0.0574 (0.0011) & 0.5598 (0.0040) & 0.2120 (0.0015) & 0.1228 (0.0011)\\
		10000 & 10 & Weibull & 1.0000 & 0.0271 (0.0003) & 0.0318 (0.0005) & 0.4848 (0.0028) & 0.1802 (0.0009) & 0.1018 (0.0006)\\
		\addlinespace
		100 & 10 & Weibull & 1.3000 & 0.0976 (0.0011) & 0.1028 (0.0019) & 0.7887 (0.0082) & 0.3099 (0.0035) & 0.1949 (0.0031)\\
		500 & 10 & Weibull & 1.3000 & 0.0464 (0.0005) & 0.0492 (0.0009) & 0.5571 (0.0042) & 0.2111 (0.0016) & 0.1230 (0.0011)\\
		1000 & 10 & Weibull & 1.3000 & 0.0332 (0.0003) & 0.0365 (0.0007) & 0.5188 (0.0033) & 0.1949 (0.0012) & 0.1118 (0.0008)\\
		10000 & 10 & Weibull & 1.3000 & 0.0165 (0.0002) & 0.0182 (0.0003) & 0.4646 (0.0027) & 0.1716 (0.0009) & 0.0960 (0.0005)\\
		\midrule
		100 & 20 & Weibull & 0.7000 & 1.3193 (0.0333) & 1.7872 (0.0444) & 1.1396 (0.0113) & 0.3491 (0.0047) & 0.2199 (0.0046)\\
		500 & 20 & Weibull & 0.7000 & 0.6388 (0.0141) & 0.9040 (0.0247) & 0.8100 (0.0096) & 0.2284 (0.0031) & 0.1220 (0.0024)\\
		1000 & 20 & Weibull & 0.7000 & 0.4429 (0.0096) & 0.7337 (0.0196) & 0.6821 (0.0072) & 0.1884 (0.0022) & 0.0938 (0.0015)\\
		10000 & 20 & Weibull & 0.7000 & 0.2364 (0.0052) & 0.6601 (0.0150) & 0.6306 (0.0048) & 0.1720 (0.0014) & 0.0823 (0.0009)\\
		\addlinespace
		100 & 20 & Weibull & 1.0000 & 0.1502 (0.0015) & 0.1592 (0.0029) & 0.9060 (0.0097) & 0.2561 (0.0032) & 0.1355 (0.0027)\\
		500 & 20 & Weibull & 1.0000 & 0.0718 (0.0007) & 0.0756 (0.0014) & 0.5919 (0.0044) & 0.1604 (0.0012) & 0.0747 (0.0007)\\
		1000 & 20 & Weibull & 1.0000 & 0.0516 (0.0005) & 0.0569 (0.0010) & 0.5283 (0.0028) & 0.1429 (0.0008) & 0.0658 (0.0005)\\
		10000 & 20 & Weibull & 1.0000 & 0.0268 (0.0002) & 0.0319 (0.0005) & 0.4740 (0.0016) & 0.1263 (0.0004) & 0.0562 (0.0003)\\
		\addlinespace
		100 & 20 & Weibull & 1.3000 & 0.0963 (0.0009) & 0.1059 (0.0019) & 0.7787 (0.0077) & 0.2153 (0.0023) & 0.1071 (0.0017)\\
		500 & 20 & Weibull & 1.3000 & 0.0457 (0.0004) & 0.0489 (0.0009) & 0.5426 (0.0032) & 0.1467 (0.0009) & 0.0677 (0.0005)\\
		1000 & 20 & Weibull & 1.3000 & 0.0327 (0.0003) & 0.0361 (0.0006) & 0.4984 (0.0022) & 0.1339 (0.0006) & 0.0609 (0.0004)\\
		10000 & 20 & Weibull & 1.3000 & 0.0162 (0.0001) & 0.0182 (0.0003) & 0.4592 (0.0014) & 0.1217 (0.0003) & 0.0531 (0.0003)\\
		\bottomrule
	\end{tabular}
\end{table}

\newpage
\begin{table}[ht!]
	\centering
	\caption{Weight-recovery summaries for Gamma model simulation study.}\label{tab-s2}
	\centering
	\fontsize{7}{11}\selectfont
	\begin{tabular}[t]{rrlllllll}
		\toprule
		n & J & distribution & shape & $W_{2,\mathrm{emp}}$ & $W_{2,\mathrm{true}}$ & $L^1$ & $L^2$ & $L^\infty$\\
		\midrule
		100 & 5 & Gamma & 0.7000 & 0.1349 (0.0014) & 0.0967 (0.0019) & 0.8880 (0.0109) & 0.4652 (0.0059) & 0.3514 (0.0054)\\
		500 & 5 & Gamma & 0.7000 & 0.0599 (0.0006) & 0.0516 (0.0010) & 0.5912 (0.0092) & 0.3090 (0.0048) & 0.2253 (0.0038)\\
		1000 & 5 & Gamma & 0.7000 & 0.0434 (0.0004) & 0.0403 (0.0007) & 0.5114 (0.0082) & 0.2663 (0.0041) & 0.1904 (0.0031)\\
		10000 & 5 & Gamma & 0.7000 & 0.0229 (0.0002) & 0.0215 (0.0003) & 0.3786 (0.0047) & 0.1978 (0.0025) & 0.1406 (0.0019)\\
		\addlinespace	
		100 & 5 & Gamma & 1.0000 & 0.1526 (0.0015) & 0.1059 (0.0022) & 0.8601 (0.0110) & 0.4520 (0.0060) & 0.3421 (0.0055)\\
		500 & 5 & Gamma & 1.0000 & 0.0687 (0.0007) & 0.0556 (0.0011) & 0.5462 (0.0081) & 0.2851 (0.0042) & 0.2061 (0.0033)\\
		1000 & 5 & Gamma & 1.0000 & 0.0496 (0.0005) & 0.0432 (0.0008) & 0.4694 (0.0070) & 0.2450 (0.0036) & 0.1741 (0.0027)\\
		10000 & 5 & Gamma & 1.0000 & 0.0261 (0.0002) & 0.0214 (0.0004) & 0.3376 (0.0040) & 0.1762 (0.0020) & 0.1254 (0.0015)\\
		\addlinespace
		100 & 5 & Gamma & 1.3000 & 0.1645 (0.0016) & 0.1173 (0.0024) & 0.7894 (0.0108) & 0.4142 (0.0058) & 0.3105 (0.0052)\\
		500 & 5 & Gamma & 1.3000 & 0.0742 (0.0007) & 0.0607 (0.0011) & 0.4962 (0.0076) & 0.2596 (0.0039) & 0.1871 (0.0030)\\
		1000 & 5 & Gamma & 1.3000 & 0.0537 (0.0005) & 0.0470 (0.0009) & 0.4275 (0.0064) & 0.2234 (0.0033) & 0.1584 (0.0024)\\
		10000 & 5 & Gamma & 1.3000 & 0.0278 (0.0002) & 0.0224 (0.0004) & 0.3084 (0.0035) & 0.1607 (0.0018) & 0.1140 (0.0014)\\
		\midrule
		100 & 10 & Gamma & 0.7000 & 0.1269 (0.0012) & 0.1089 (0.0020) & 1.0259 (0.0110) & 0.4159 (0.0051) & 0.2832 (0.0049)\\
		500 & 10 & Gamma & 0.7000 & 0.0586 (0.0006) & 0.0546 (0.0010) & 0.7195 (0.0077) & 0.2759 (0.0030) & 0.1644 (0.0023)\\
		1000 & 10 & Gamma & 0.7000 & 0.0424 (0.0004) & 0.0417 (0.0007) & 0.6456 (0.0062) & 0.2463 (0.0024) & 0.1447 (0.0017)\\
		10000 & 10 & Gamma & 0.7000 & 0.0225 (0.0002) & 0.0224 (0.0003) & 0.5383 (0.0034) & 0.2034 (0.0012) & 0.1186 (0.0009)\\
		\addlinespace
		100 & 10 & Gamma & 1.0000 & 0.1455 (0.0013) & 0.1189 (0.0023) & 0.9839 (0.0103) & 0.3924 (0.0047) & 0.2599 (0.0044)\\
		500 & 10 & Gamma & 1.0000 & 0.0674 (0.0006) & 0.0585 (0.0011) & 0.6875 (0.0068) & 0.2620 (0.0026) & 0.1535 (0.0018)\\
		1000 & 10 & Gamma & 1.0000 & 0.0490 (0.0004) & 0.0442 (0.0008) & 0.6181 (0.0053) & 0.2349 (0.0020) & 0.1368 (0.0014)\\
		10000 & 10 & Gamma & 1.0000 & 0.0256 (0.0002) & 0.0223 (0.0004) & 0.5234 (0.0031) & 0.1961 (0.0010) & 0.1126 (0.0008)\\
		\addlinespace
		100 & 10 & Gamma & 1.3000 & 0.1569 (0.0014) & 0.1317 (0.0025) & 0.9187 (0.0101) & 0.3634 (0.0044) & 0.2355 (0.0039)\\
		500 & 10 & Gamma & 1.3000 & 0.0728 (0.0007) & 0.0641 (0.0012) & 0.6484 (0.0063) & 0.2462 (0.0024) & 0.1434 (0.0016)\\
		1000 & 10 & Gamma & 1.3000 & 0.0530 (0.0005) & 0.0481 (0.0009) & 0.5876 (0.0047) & 0.2223 (0.0018) & 0.1285 (0.0012)\\
		10000 & 10 & Gamma & 1.3000 & 0.0274 (0.0002) & 0.0234 (0.0004) & 0.5062 (0.0028) & 0.1884 (0.0009) & 0.1073 (0.0007)\\
		\midrule
		100 & 20 & Gamma & 0.7000 & 0.1242 (0.0011) & 0.1105 (0.0021) & 1.0165 (0.0110) & 0.2947 (0.0039) & 0.1654 (0.0033)\\
		500 & 20 & Gamma & 0.7000 & 0.0580 (0.0005) & 0.0549 (0.0010) & 0.6861 (0.0073) & 0.1866 (0.0020) & 0.0881 (0.0012)\\
		1000 & 20 & Gamma & 0.7000 & 0.0422 (0.0004) & 0.0418 (0.0007) & 0.6031 (0.0052) & 0.1640 (0.0014) & 0.0761 (0.0008)\\
		10000 & 20 & Gamma & 0.7000 & 0.0225 (0.0002) & 0.0224 (0.0003) & 0.5192 (0.0024) & 0.1400 (0.0006) & 0.0636 (0.0004)\\
		\addlinespace
		100 & 20 & Gamma & 1.0000 & 0.1434 (0.0012) & 0.1200 (0.0024) & 0.9562 (0.0104) & 0.2708 (0.0034) & 0.1446 (0.0027)\\
		500 & 20 & Gamma & 1.0000 & 0.0670 (0.0006) & 0.0584 (0.0011) & 0.6499 (0.0063) & 0.1762 (0.0017) & 0.0822 (0.0010)\\
		1000 & 20 & Gamma & 1.0000 & 0.0488 (0.0004) & 0.0442 (0.0008) & 0.5773 (0.0042) & 0.1564 (0.0012) & 0.0717 (0.0007)\\
		10000 & 20 & Gamma & 1.0000 & 0.0257 (0.0002) & 0.0222 (0.0004) & 0.5023 (0.0020) & 0.1350 (0.0005) & 0.0608 (0.0004)\\
		\addlinespace
		100 & 20 & Gamma & 1.3000 & 0.1547 (0.0013) & 0.1322 (0.0025) & 0.8868 (0.0100) & 0.2487 (0.0032) & 0.1293 (0.0024)\\
		500 & 20 & Gamma & 1.3000 & 0.0724 (0.0006) & 0.0641 (0.0012) & 0.6122 (0.0055) & 0.1656 (0.0015) & 0.0762 (0.0009)\\
		1000 & 20 & Gamma & 1.3000 & 0.0528 (0.0004) & 0.0481 (0.0008) & 0.5482 (0.0036) & 0.1483 (0.0010) & 0.0676 (0.0006)\\
		10000 & 20 & Gamma & 1.3000 & 0.0274 (0.0002) & 0.0234 (0.0004) & 0.4872 (0.0018) & 0.1303 (0.0005) & 0.0579 (0.0003)\\
		\bottomrule
	\end{tabular} 
\end{table}

\newpage

\begin{table}[ht!]
	\centering
	\caption{Weight-recovery summaries for mixed model simulation study.}\label{tab-s3}
	\centering
	\fontsize{7}{11}\selectfont
	\begin{tabular}[t]{rrlllllll}
		\toprule
		n & J & distribution & shape & $W_{2,\mathrm{emp}}$ & $W_{2,\mathrm{true}}$ & $L^1$ & $L^2$ & $L^\infty$\\
		\midrule
		100 & 5 & Mixed & 0.7000 & 0.5502 (0.0243) & 0.6033 (0.0308) & 0.8172 (0.0117) & 0.4352 (0.0066) & 0.3327 (0.0060)\\
		500 & 5 & Mixed & 0.7000 & 0.2497 (0.0113) & 0.2689 (0.0136) & 0.4951 (0.0089) & 0.2623 (0.0048) & 0.1936 (0.0040)\\
		1000 & 5 & Mixed & 0.7000 & 0.1882 (0.0087) & 0.2412 (0.0133) & 0.4053 (0.0072) & 0.2146 (0.0039) & 0.1576 (0.0032)\\
		10000 & 5 & Mixed & 0.7000 & 0.1076 (0.0049) & 0.1955 (0.0105) & 0.3178 (0.0049) & 0.1669 (0.0026) & 0.1215 (0.0022)\\
		\addlinespace
		100 & 5 & Mixed & 1.0000 & 0.1479 (0.0014) & 0.1225 (0.0024) & 0.8448 (0.0112) & 0.4460 (0.0062) & 0.3394 (0.0058)\\
		500 & 5 & Mixed & 1.0000 & 0.0679 (0.0006) & 0.0654 (0.0012) & 0.5220 (0.0077) & 0.2706 (0.0040) & 0.1935 (0.0032)\\
		1000 & 5 & Mixed & 1.0000 & 0.0499 (0.0005) & 0.0499 (0.0009) & 0.4186 (0.0057) & 0.2192 (0.0030) & 0.1568 (0.0023)\\
		10000 & 5 & Mixed & 1.0000 & 0.0251 (0.0002) & 0.0265 (0.0004) & 0.2995 (0.0033) & 0.1548 (0.0016) & 0.1079 (0.0012)\\
		\addlinespace
		100 & 5 & Mixed & 1.3000 & 0.1329 (0.0013) & 0.1246 (0.0024) & 0.7077 (0.0099) & 0.3697 (0.0053) & 0.2726 (0.0046)\\
		500 & 5 & Mixed & 1.3000 & 0.0627 (0.0006) & 0.0618 (0.0011) & 0.4226 (0.0065) & 0.2204 (0.0034) & 0.1579 (0.0026)\\
		1000 & 5 & Mixed & 1.3000 & 0.0460 (0.0004) & 0.0469 (0.0008) & 0.3388 (0.0048) & 0.1768 (0.0024) & 0.1257 (0.0018)\\
		10000 & 5 & Mixed & 1.3000 & 0.0229 (0.0002) & 0.0231 (0.0004) & 0.2467 (0.0027) & 0.1292 (0.0014) & 0.0922 (0.0010)\\
		\midrule
		100 & 10 & Mixed & 0.7000 & 0.5601 (0.0251) & 0.6829 (0.0304) & 0.9574 (0.0114) & 0.3961 (0.0057) & 0.2776 (0.0057)\\
		500 & 10 & Mixed & 0.7000 & 0.2830 (0.0107) & 0.3709 (0.0187) & 0.6814 (0.0082) & 0.2691 (0.0037) & 0.1744 (0.0034)\\
		1000 & 10 & Mixed & 0.7000 & 0.2138 (0.0086) & 0.3189 (0.0165) & 0.6221 (0.0069) & 0.2420 (0.0030) & 0.1522 (0.0027)\\
		10000 & 10 & Mixed & 0.7000 & 0.1204 (0.0048) & 0.2526 (0.0110) & 0.5738 (0.0053) & 0.2187 (0.0022) & 0.1349 (0.0019)\\
		\addlinespace
		100 & 10 & Mixed & 1.0000 & 0.1431 (0.0014) & 0.1380 (0.0025) & 0.9349 (0.0099) & 0.3750 (0.0045) & 0.2482 (0.0043)\\
		500 & 10 & Mixed & 1.0000 & 0.0665 (0.0006) & 0.0676 (0.0013) & 0.6264 (0.0055) & 0.2392 (0.0022) & 0.1419 (0.0017)\\
		1000 & 10 & Mixed & 1.0000 & 0.0488 (0.0005) & 0.0521 (0.0009) & 0.5589 (0.0038) & 0.2119 (0.0015) & 0.1234 (0.0011)\\
		10000 & 10 & Mixed & 1.0000 & 0.0247 (0.0002) & 0.0277 (0.0004) & 0.4951 (0.0025) & 0.1855 (0.0008) & 0.1050 (0.0006)\\
		\addlinespace
		100 & 10 & Mixed & 1.3000 & 0.1257 (0.0012) & 0.1297 (0.0024) & 0.7874 (0.0083) & 0.3072 (0.0035) & 0.1923 (0.0030)\\
		500 & 10 & Mixed & 1.3000 & 0.0594 (0.0005) & 0.0602 (0.0011) & 0.5612 (0.0037) & 0.2108 (0.0014) & 0.1213 (0.0011)\\
		1000 & 10 & Mixed & 1.3000 & 0.0437 (0.0004) & 0.0459 (0.0008) & 0.5228 (0.0028) & 0.1951 (0.0010) & 0.1116 (0.0008)\\
		10000 & 10 & Mixed & 1.3000 & 0.0219 (0.0002) & 0.0227 (0.0003) & 0.4921 (0.0022) & 0.1801 (0.0007) & 0.0992 (0.0005)\\
		\midrule
		100 & 20 & Mixed & 0.7000 & 0.5220 (0.0149) & 0.7279 (0.0216) & 0.9271 (0.0102) & 0.2759 (0.0039) & 0.1645 (0.0036)\\
		500 & 20 & Mixed & 0.7000 & 0.2724 (0.0077) & 0.3974 (0.0143) & 0.6471 (0.0062) & 0.1826 (0.0021) & 0.0958 (0.0018)\\
		1000 & 20 & Mixed & 0.7000 & 0.2038 (0.0060) & 0.3089 (0.0103) & 0.5876 (0.0046) & 0.1637 (0.0015) & 0.0829 (0.0013)\\
		10000 & 20 & Mixed & 0.7000 & 0.1084 (0.0032) & 0.2694 (0.0082) & 0.5426 (0.0029) & 0.1489 (0.0009) & 0.0730 (0.0008)\\
		\addlinespace
		100 & 20 & Mixed & 1.0000 & 0.1398 (0.0013) & 0.1411 (0.0025) & 0.9005 (0.0100) & 0.2603 (0.0035) & 0.1455 (0.0030)\\
		500 & 20 & Mixed & 1.0000 & 0.0664 (0.0006) & 0.0699 (0.0012) & 0.5956 (0.0042) & 0.1611 (0.0012) & 0.0751 (0.0009)\\
		1000 & 20 & Mixed & 1.0000 & 0.0489 (0.0004) & 0.0525 (0.0009) & 0.5373 (0.0028) & 0.1440 (0.0008) & 0.0649 (0.0005)\\
		10000 & 20 & Mixed & 1.0000 & 0.0245 (0.0002) & 0.0282 (0.0004) & 0.4795 (0.0015) & 0.1267 (0.0003) & 0.0543 (0.0003)\\
		\addlinespace
		100 & 20 & Mixed & 1.3000 & 0.1238 (0.0011) & 0.1309 (0.0024) & 0.7509 (0.0076) & 0.2088 (0.0023) & 0.1035 (0.0017)\\
		500 & 20 & Mixed & 1.3000 & 0.0595 (0.0005) & 0.0623 (0.0011) & 0.5385 (0.0027) & 0.1454 (0.0008) & 0.0653 (0.0005)\\
		1000 & 20 & Mixed & 1.3000 & 0.0439 (0.0004) & 0.0464 (0.0008) & 0.5003 (0.0017) & 0.1345 (0.0005) & 0.0595 (0.0004)\\
		10000 & 20 & Mixed & 1.3000 & 0.0221 (0.0002) & 0.0230 (0.0003) & 0.4674 (0.0011) & 0.1237 (0.0002) & 0.0528 (0.0002)\\
		\bottomrule
	\end{tabular}
\end{table}

\newpage
\section{Supplementary material from Section \ref{sec-5}}

\begin{table}[ht]
	\centering
		\caption{Training sites with WGS84 geographical coordinates}\label{training_sites}
	\fontsize{9}{11}\selectfont
	\begin{tabular}{rlccrlcc}
		\toprule
		\# & \textbf{Site} & \textbf{Lat (°N)} & \textbf{Lon (°E)} &
		\# & \textbf{Site} & \textbf{Lat (°N)} & \textbf{Lon (°E)} \\
		\midrule
		1  & Agia              & 39.731 & 22.758 & 19 & Nea Anchialos     & 39.282 & 22.818 \\
		2  & Agiokampos        & 39.757 & 22.760 & 20 & Nea Ionia          & 39.385 & 22.906 \\
		3  & Agrafa            & 39.106 & 21.622 & 21 & Neochori Trikalon  & 39.400 & 21.660 \\
		4  & Almyros           & 39.184 & 22.757 & 22 & Palaiokastro       & 39.588 & 22.554 \\
		5  & Ampelonas         & 39.758 & 22.382 & 23 & Palaiopyrgos       & 39.609 & 22.540 \\
		6  & Ampelos           & 39.638 & 22.477 & 24 & Palamas            & 39.464 & 22.081 \\
		7  & Armenio           & 39.493 & 22.612 & 25 & Platykampos        & 39.513 & 22.482 \\
		8  & Domokos           & 39.129 & 22.328 & 26 & Pteleos            & 39.010 & 22.872 \\
		9  & Elassona          & 39.893 & 22.188 & 27 & Pyli               & 39.457 & 21.627 \\
		10 & Farkadona         & 39.581 & 22.354 & 28 & Sofades            & 39.326 & 22.188 \\
		11 & Farsala           & 39.299 & 22.386 & 29 & Stylida            & 38.920 & 22.628 \\
		12 & Kalampaka         & 39.710 & 21.629 & 30 & Sykia              & 39.548 & 22.436 \\
		13 & Karditsa          & 39.367 & 21.921 & 31 & Trikala            & 39.555 & 21.768 \\
		14 & Karitsa           & 39.668 & 22.827 & 32 & Tyrnavos           & 39.739 & 22.290 \\
		15 & Krania            & 39.962 & 22.050 & 33 & Velestino          & 39.396 & 22.744 \\
		16 & Larissa           & 39.643 & 22.413 & 34 & Volos              & 39.373 & 22.943 \\
		17 & Makrinitsa        & 39.386 & 22.999 & 35 & Zagora             & 39.450 & 23.091 \\
		18 & Mouzaki           & 39.441 & 21.683 &     &                   &        &        \\
		\bottomrule
	\end{tabular}
\end{table}

\begin{table}[ht!]
	\centering
		\caption{Test sites with WGS84 coordinates and brief climatic characteristics.}
	\fontsize{9}{11}\selectfont
	\begin{tabular}{rlccc}
		\toprule
		\# & \textbf{Site} & \textbf{Lat (°N)} & \textbf{Lon (°E)} & \textbf{Microclimatic and Spatial Features} \\
		\midrule
		1.  &Eastern\_Coast         & 39.350 & 23.150 & Slight extrapolation beyond Volos \\
		2.  &Farsala\_Plain           & 39.170 & 22.240 & Inland plain; high summer maxima \\
		3.  &Lamia\_Corridor       & 39.120 & 21.950 & Southern inland pocket; hot continental climate \\
		4.  &Larissa\_Volos\_Corridor & 39.260 & 22.620 & Interpolation region; transport corridor \\
		5.  &Nea\_Anchialos\_Coast & 39.330 & 22.900 & Coastal site; maritime influence \\
		6.  &Pteleos\_Coast          & 39.020 & 22.600 & Eastern coastal plain \\
		7.  &South\_Larissa           & 39.060 & 22.320 & Between Larissa and Farsala; lowland corridor \\
		8.  &Syllos                          & 39.230 & 22.150 & Southern Thessalian plain; warm inland basin \\
		9.  &Tyrnavos\_Outskirts  & 39.650 & 22.450 & Agricultural plain north of Larissa \\
		10. &Velestino\_Corridor   & 39.480 & 22.780 & Lowland corridor toward Volos \\
		\bottomrule
	\end{tabular}
	\label{tab:unknown_sites}
\end{table}

\begin{figure}[ht!]
	\centering
	\includegraphics[width=5.5in]{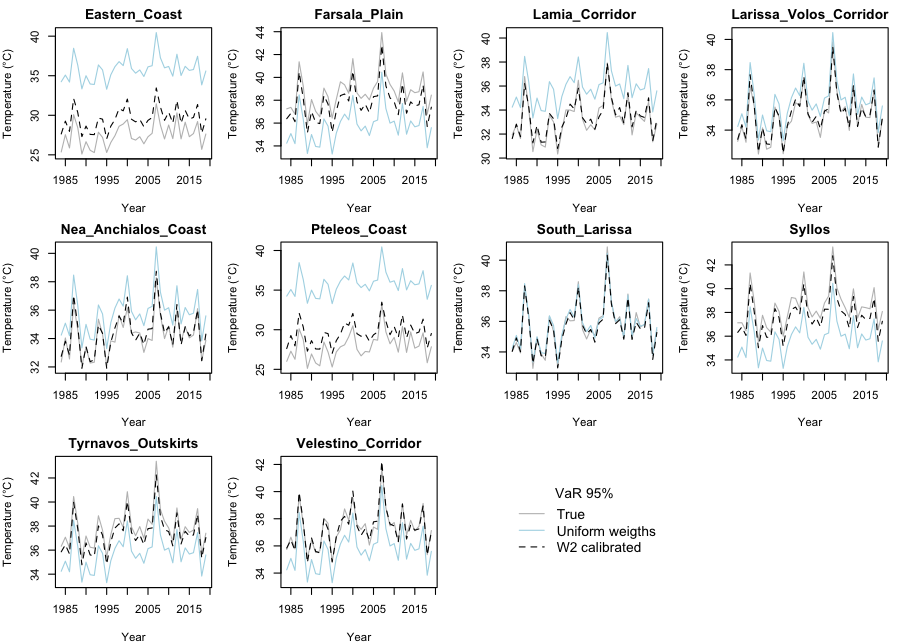} 
	\caption{Illustration of the estimation for the 95\% VaR by the uniformly weighted barycenter (light blue line) and the calibrated barycenter (black dashed line) averaging models.}\label{VaR_0950}
\end{figure}

\begin{figure}[ht!]
	\centering
	\includegraphics[width=5.5in]{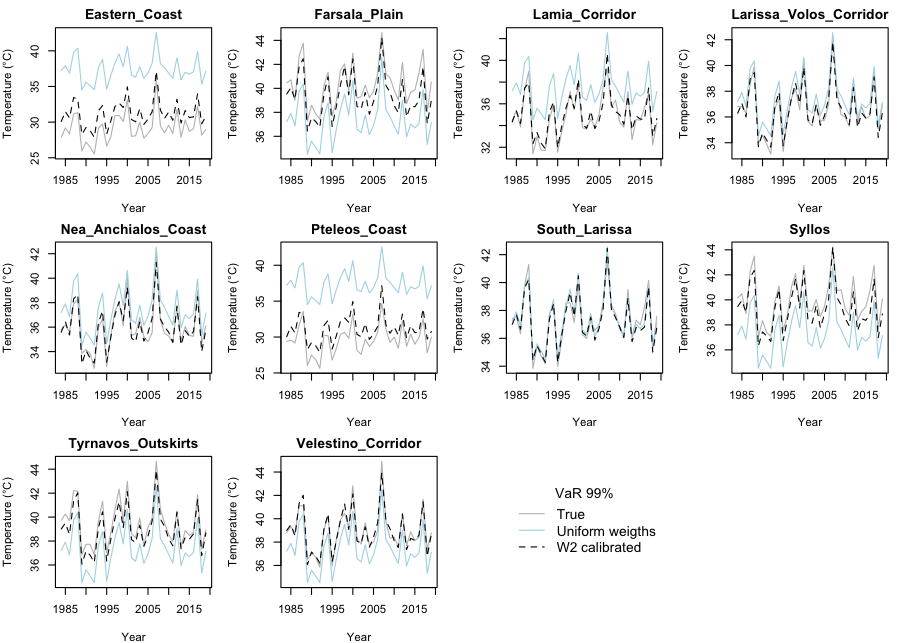} 
	\caption{Illustration of the estimation for the 99\% VaR by the uniformly weighted barycenter (light blue line) and the calibrated barycenter (black dashed line) averaging models.}\label{VaR_0990}
\end{figure}

\end{document}